\documentclass[twocolumn,prd,superscriptaddress, colorlinks, linkcolor=refcol, citecolor=refcol, urlcolor=refcol]{revtex4-2}
\pdfoutput=1
\usepackage{bm}                            
\usepackage{graphicx}                                          
\usepackage{booktabs}                                                 \usepackage{amssymb,amsmath,amsfonts}                                 
\usepackage{soul}
\usepackage{orcidlink}             
\usepackage[utf8]{inputenc}
\usepackage{slashed}
\usepackage{cleveref}
\usepackage{graphicx}
\usepackage{hyperref}
\usepackage{amssymb,amsmath,amsfonts}
\usepackage{color}
\usepackage[utf8]{inputenc}
\usepackage[caption=false]{subfig}

\newcommand{\dr}{\ensuremath{\mathrm{d}}}

\newcommand{\photonSelfE}{\ensuremath{\Pi}}

\newcommand{\cO}{\ensuremath{\mathcal{O}}}
\newcommand{\cL}{\ensuremath{\mathcal{L}}}

\def \beq  {\begin{equation}}
\def \eeq  {\end{equation}}
\def \beqar {\begin{eqnarray}}
\def \eeqar {\end{eqnarray}}
\def\sqr#1#2{{\vcenter{\vbox{\hrule height.#2pt
\hbox{\vrule width.#2pt height#1pt \kern#1pt
\vrule width.#2pt}\hrule height.#2pt}}}}

\definecolor{refcol}{RGB}{51,80,182}
\usepackage{microtype}
\definecolor{red}{rgb}{1,0,0}

\newcommand{\Tr}{\text{Tr\,}}

\usepackage{ulem}

\begin{document}

\begin{raggedright}
%% DO NOT COMMMENT OUT
    \hfill MIT-CTP 5773, LA-UR-24-30156\\
\end{raggedright}

\title{Dilepton production from moaton quasiparticles}

\author{Zohar Nussinov \, \orcidlink{0000-0001-7947-7829} }
\email{zohar@wustl.edu}
\affiliation{Physics Department, Washington University, St. Louis, MO 63130, USA}
\affiliation{Rudolf Peierls Centre for Theoretical Physics, University of Oxford, Oxford OX1 3PU, United Kingdom}
\author{Michael C. Ogilvie \, \orcidlink{0000-0002-3586-5237}}
\email{mco@wustl.edu}
\affiliation{Physics Department, Washington University, St. Louis, MO 63130, USA}
\author{Laurin Pannullo \, \orcidlink{0000-0003-4808-5819} }
\email{lpannullo@physik.uni-bielefeld.de}
\affiliation{Fakultät für Physik, Universität Bielefeld, 
%Universitätsstraße 25, 
D-33615 Bielefeld, Germany}
\author{ Robert D. Pisarski \,\orcidlink{0000-0002-7862-4759} }
\email{pisarski@bnl.gov}
\affiliation{Department of Physics, Brookhaven National Laboratory, Upton, NY 11973}
\author{Fabian Rennecke \, \orcidlink{0000-0003-1448-677X}}
\email{fabian.rennecke@theo.physik.uni-giessen.de}
\affiliation{Institute for Theoretical Physics, Justus Liebig University Giessen, %Heinrich-Buff-Ring 16, 
35392 Giessen, Germany}
\affiliation{Helmholtz Research Academy Hesse for FAIR (HFHF), Campus Giessen, Giessen, Germany}
\author{Stella T.~Schindler \, \orcidlink{0000-0003-3224-4627} }
\email{schindler@lanl.gov}
\affiliation{Theoretical Division, Los Alamos National Laboratory, Los Alamos, NM 87545, USA}
\affiliation{Center for Theoretical Physics, Massachusetts Institute of Technology, 
%77 Massachusetts Ave., 
Cambridge, MA 02139, USA}
\author{Marc Winstel  \, \orcidlink{0000-0003-3470-9926}}
\email{winstel@itp.uni-frankfurt.de}
\thanks{corresponding author}
\affiliation{Institut f{\"u}r Theoretische Physik, J. W. Goethe-Universit{\"a}t, %Frankfurt, Max-von-Laue Stra{\ss}e 1, 
D-60438 Frankfurt am Main, Germany}

\begin{abstract}
The phase diagram of QCD probably exhibits a moat regime over a large region of temperature $T$ and chemical potential $\mu\neq0$. A moat regime is characterized by quasiparticle \textit{moatons} (pions) whose energy is  minimal at nonzero spatial momentum. At $\mu\neq 0$, higher mass dimension operators play a critical role in a moat regime. At dimension six, there are nine possible gauge invariant couplings between scalars and photons. For back-to-back dilepton production, only one operator contributes, which significantly enhances production near a moat threshold. This enhancement is an experimental signature of moatons. 
\end{abstract}

\maketitle

The phase diagram of Quantum Chromodynamics (QCD) at a nonzero temperature $T$ and quark chemical potential $\mu$ (or baryon chemical potential $\mu_B = 3 \mu$), is important for a variety of physical systems, from neutron stars, to heavy ion collisions, to the early universe. 
We know much about the region of $T \neq 0$ at low $\mu \ll T$ from heavy ion colliders operating at ultra-relativistic energies, such as the Relativistic Heavy Ion Collider (RHIC) at BNL and the Large Hadron Collider at CERN.
We also have a wealth of understanding of $\mu=0$ behavior from first-principles lattice QCD techniques, which indicate that the chiral transition is crossover at a temperature $T \approx 156.5 \pm 1.5$~MeV
\cite{HotQCD:2018pds,Borsanyi:2020fev}.
However, the lattice cannot reach beyond $\mu \sim T$ on classical computers due to  
the sign problem. 
A first-principles understanding of heavy ion collisions at moderate energies requires studying QCD at $\mu > T$, and neutron stars 
have large $\mu \gg T$.  

As $\mu$ increases, it is likely that a Critical End Point (CEP) develops, where a line of crossover
transitions meets a line of first order transitions
\cite{Asakawa:1989bq,Stephanov:1998dy,Stephanov:1999zu,Parotto:2018pwx,Bzdak:2019pkr}. From direct computations in QCD using the
Functional Renormalization Group (FRG) \cite{Fu:2019hdw} and
Schwinger-Dyson equations \cite{Gao:2020qsj, Gao:2020fbl, Gunkel:2021oya}, as well as reconstructions based on lattice data
\cite{Basar:2023nkp, Clarke:2024ugt}, this is estimated to occur
for $(T_\chi,\mu_\chi) \approx (100,200)$~MeV.  This lies in the region $\sqrt{s}/A: 3 \rightarrow 7$~GeV
in heavy ion collisions.  By perverse coincidence, the Beam Energy Scan by the STAR experiment at RHIC studied collisions down to $\sqrt{s}/A = 7$~GeV, and at $\sqrt{s}/A = 3$~GeV in fixed target mode, but not in between.  
While intriguing, their results did not find evidence for a CEP.  This region will be studied at the Nuclotron-based Ion Collider Facility
(NICA) at the Joint Institute for Nuclear Research, and 
with the Compressed Baryon Matter (CBM) experiment at 
the Facility for Antiproton and Ion Research (FAIR) at the GSI Helmholtz Center for Heavy Ion Research.

The CEP, while of fundamental significance, may be difficult to probe directly.  In vacuum, the mass of the $\sigma$ meson is large, perhaps $\sim\! 500$~MeV, 
with a width of similar order, 
rendering it difficult to extract experimentally. 
In contrast, at the CEP the $\sigma$ is precisely massless \footnote{To be more precise, the critical mode is a mixture between the $\sigma$, the $\omega^0$ meson and the Polyakov loops, and the relevant mass that vanishes at the CEP is the spacelike screening mass \cite{Haensch:2023sig}.} but pions remain massive.  Because the $\sigma$'s singularity is so far from the origin in vacuum, it must travel a long way, suggesting that the basin of attraction to the CEP can be narrow.
This naive argument is confirmed by 
studies of the chiral phase transition, where critical scaling is absent even 
close to the transition; see Refs.
\cite{Braun:2010vd, Schaefer:2011ex, Fu:2019hdw, Braun:2020ada, Gao:2021vsf, Fu:2023lcm, Bernhardt:2023hpr, Braun:2023qak, Lu:2023mkn,Fu:2024rto}. 

A compelling target for determining where novel phenomena occur is a ``moat'' regime, which occurs when the dispersion relations of quark correlations
in meson channels have a global minimum at {\it non}zero spatial momentum.
The 
FRG \cite{Fu:2019hdw,Fu:2024rto} indicates  
that the moat regime in QCD is
 {\it extensive}, 
occurring over a {\it large} region of the phase diagram at finite $T$ and $\mu$, 
including the vicinity of the CEP. 
As detailed below, a moat regime gives rise to
exotic behavior involving spatial modulations, which may (but need not) include
a region with inhomogeneous phases.

In this Letter we propose a striking signature of moat regimes: new
thresholds with a significant enhancement in dilepton production in heavy ion collisions. Experimental observation of a moat regime would confirm the existence of
novel phases in QCD, and help locate the CEP.\\

\begin{figure}[t]
    \centering
    \includegraphics[trim=0cm 0.4cm 0cm 0.3cm,
    clip,
    keepaspectratio,
    width=.9\columnwidth]{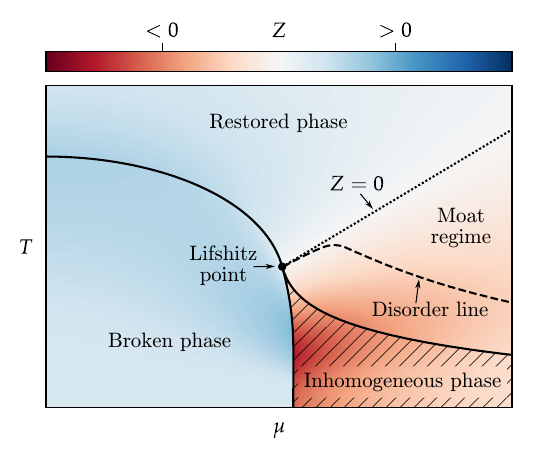}
    \caption{
    A phase diagram sketch based on the Gross-Neveu model in Reference~\cite{Koenigstein:2021llr}, which exhibits \textit{all} three characteristic behaviors of moat regimes: a disorder line, oscillatory regime, and inhomogeneous phase.  Red shading indicates a moat regime, where $Z\!<\!0$. Solid lines separate different phases. Dashed lines indicate qualitative changes in system correlations, related to spatial modulations; the oscillatory regime lies between the disorder line and the inhomogeneous phase.
    The exact location and nature of each transition is model dependent. However, an extensive moat regime is a common feature of models, and is also seen in FRG analyses of QCD \cite{Fu:2019hdw}.
    In the Gross-Neveu model, the CEP is replaced by a Lifshitz point due to the emergence of an inhomogeneous phase.
    }
     \label{fig:sketch}
\end{figure}

\noindent {\bf Moat regimes at $\mu \neq 0$.} 
A moat regime is characterized by a dispersion relation exhibiting a local minimum at nonzero momentum, which can induce exotic phase behaviors.
Often, moat regimes onset when the wave function renormalization constant $Z$ for the quadratic spatial momentum term becomes negative.
This typically gives rise to a ``disorder line" in the phase diagram
\cite{stephenson_ising_1970,Schindler:2019ugo,Schindler:2021otf}, 
separating a region in which the two-point correlation function has exponential falloff at large distances, from a region in which
it behaves as an exponential times an oscillatory function;
this ``oscillatory regime'' is similar to Friedel oscillations, which are screened at non-zero temperature \cite{fetter2012quantum}.
If the dip in the moat dispersion relation
becomes sufficiently deep, then the
moat regime may additionally
develop a finite-momentum
(i.e., spatially-nonuniform) condensate, such as, e.g., pion, kaon, or superconducting condensates \cite{Overhauser:1960,Migdal:1971,Sawyer:1972cq,Migdal:1973zm,Kaplan:1986yq, Kryjevski:2005qq, Schaefer:2006ds}.
Direct transitions
from a spatially homogeneous broken symmetry phase to 
an inhomogeneous phase may also arise (see Fig.\ \ref{fig:sketch}).

A natural question is whether 
a moat regime exists in QCD-like theories or
is an artifact of a given approximation method.
Moat regimes  are {\it ubiquitous}
in condensed matter, which like finite-density QCD breaks Lorentz invariance;
see Refs.~\cite{Yoshimori,Hornnerich,Seul,modulation1,modulation2} 
and \footnote{This also includes  
models of magnetic materials \cite{Seul,Yoshimori,Hornnerich,Selke}, elastic stressed systems \cite{Amnon}, liquid crystals \cite{Elihu,LC2}, chemical mixtures and membranes \cite{Seul}, and general Ginzburg-Landau theories describing competing 
orders \cite{compete_GL}. In these and numerous other arenas, the interplay between the varying order gradient terms leads to low energy states which may obtain their minima at finite wave-vectors. The resulting moat-type structures underlie basic aspects of crystallization \cite{Braz,Braz2}, superconductors in external magnetic fields \cite{FFLO1,FFLO2,FFLO3}, quantum Hall systems \cite{qhstr2,QHE,fogler1,chalker,qhstr1, qhstr3,Pu2024}, and countless others.}.
Indeed, the term ``moat'' hails from condensed matter \cite{Pu2024,Sedrakyan14}.
Furthermore, moat regimes naturally arise
in theories with combined charge and complex conjugation ($\mathcal{CK}$) symmetry \cite{Schindler:2019ugo}.
Because the Dirac operator is non-Hermitian at $\mu \neq 0$,
QCD has $\mathcal{CK}$ symmetry \cite{Meisinger:2012va, Schindler:2021otf}.
This is a particular case of
broad non-Hermitian
theories with generalized $\mathcal{PT}$ 
symmetries~\cite{Bender:1998ke,Bender:2019cwm, Bender:2023cem}.

Moat regimes and their associated phases appear in many models sharing features with QCD. A canonical example is the Gross-Neveu model in $(1+1)$ dimensions. 
This theory is asymptotically free, spontaneously breaks chiral symmetry
through dynamical generation of mass, and is tractable when
the number of fermions $N \rightarrow \infty$.
The phase diagram at low $\mu < T$ is similar to what we suspect occurs in QCD in the chiral limit: 
a line of second order transitions extends
from $\mu = 0$ and $T \neq 0$.
Due to the emergence of the Peierls transition, 
this line ends not in a CEP but in a Lifshitz point; for lower $T$, there is a region with spatially inhomogeneous condensates \cite{Thies:2003kk,Basar:2008ki,Basar:2009fg,Lenz:2020bxk,Lenz:2021kzo, Lenz:2021vdz, Nonaka:2021pwm,Pannullo:2023cat,Koenigstein:2023yzv}.  
Crucially, the Gross-Neveu model has a moat regime
in a broad region of the $T$-$\mu$ plane \cite{Koenigstein:2021llr} 
(Fig.\ \ref{fig:sketch}).

Numerous studies of moat regimes have been carried out in (3+1)-dimensional models sharing features of QCD.
In the Nambu-Jona-Lasino (NJL) model \cite{Nickel:2009wj,Buballa:2015awa,Seul,Lee:2015bva,Hidaka:2015xza,Buballa:2018hux,Carignano:2019ivp}, the presence of an inhomogeneous phase depends strongly upon the regularization scheme, while a moat regime is robust \cite{Pannullo:2024sov}.
Schwinger-Dyson analysis shows that the symmetric solution is unstable to an inhomogeneous phase at low $T$ and high $\mu$ \cite{Motta:2024agi,Motta:2024rvk}. 
Disorder lines 
appear in Polyakov-Nambu-Jona-Lasino (PNJL) models \cite{Nishimura:2014rxa, Nishimura:2014kla}, Polyakov-quark-meson models \cite{Haensch:2023sig}, quark-meson models \cite{Topfel:2024iop}, static quark models at strong coupling \cite{Nishimura:2015lit}, liquid-gas models \cite{Nishimura:2016yue}, and in four-fermion models in $(2+1)$ dimensions \cite{Winstel:2024dqu} \footnote{In perturbative QCD with massless quarks, Friedel oscillations arise as an oscillatory function times a power law fall-off.\cite{Kapusta:1988fi}}. 
For scalar models with global $O(N)$ symmetry, Goldstone bosons can disorder the system into  a quantum pion liquid \cite{Pisarski:2018bct, Pisarski:2020dnx, Winstel:2024qle}. 
Moat regimes are more prevalent in models of hadronic physics at nonzero density than inhomogeneous phases. 

Refs.\ \cite{Pisarski:2020gkx, Pisarski:2021qof} 
showed that a moat regime affects
particle production through 
two-particle correlations generated by thermodynamic
fluctuations \cite{Pisarski:2021qof} and Hanbury 
Brown-Twiss interferometry for pions \cite{Rennecke:2023xhc, Fukushima:2023tpv}.
In this Letter, we compute how quasiparticle pions in a moat regime,
which we  
dub ``moatons'', affect dilepton production.  For technical reasons, we limit ourselves to 
dileption
pairs with zero spatial momentum, and leave the  
analysis of nonzero spatial momentum
to a later publication.  

Related analyses of dilepton production have been carried out
in a color superconductor \cite{Nishimura:2022mku}, 
near a critical endpoint \cite{Nishimura:2023oqn,Nishimura:2024kvz}, and in a chiral spiral
~\cite{Hayashi:2024sae}
and 
\footnote{
As this work was in preparation, Ref.~\cite{Hayashi:2024sae} appeared, and analyzed
dilepton production from a specific inhomogeneous phase.
Ref. \cite{Hayashi:2024sae} includes the modified dispersion relation
of a dual chiral density wave, but not the changes to the photon-moaton
vertices.  This is discussed
in the Supplemental Material.
}.
\\

\noindent {\bf Effective Lagrangian for the moat regime.} 
A simple model of a moat regime is given by the Lagrangian \cite{Pisarski:2018bct,Pisarski:2020dnx}
\begin{align}\label{eff_lag}  
  {\cal L} = &\frac{1}{2} (\partial_0 \vec{\phi})^2 
  + \frac{Z}{2} (\partial_i \vec{\phi})^2 
   + \frac{1}{2 M^2}(\partial_i^2 \vec{\phi})^2 \nonumber\\
   &\!+\! \frac{m^2}{2} \vec{\phi}^{\, 2}+ \frac{\lambda}{8}
   (\vec{\phi}^{\,2})^2 \; ,
\end{align}
where $\vec{\phi}=(\sigma,\vec{\pi})$.
This Lagrangian is written in the preferred frame provided by the thermal bath, using a Euclidean metric.  To maintain
causality, we insist that time derivatives only enter quadratically.

A moat regime arises in Eq.~\eqref{eff_lag}
when $Z$ is negative, so the energy has a minimum at non-zero momentum. This was first observed in QCD at finite $T$ and $\mu$ in an FRG analysis \cite{Fu:2019hdw}.
Consequently, we must include a non-renormalizable term
quartic in spatial derivatives in Eq.~\eqref{eff_lag}.
While this term may look unfamiliar,
it is
natural in derivative expansions of effective Lagrangians. 
Examples include systems near second-order phase transitions \cite{Balog:2019rrg} and 
condensed matter systems  \cite{Yoshimori,Hornnerich,Seul,modulation1,modulation2}
(where such derivatives may
be replaced by their discrete lattice counterparts \footnote{On a lattice, the continuum limit $k^4$ term is replaced by a squared lattice Laplacian corresponding to competing nearest and next nearest neighbor interactions.}).

{\it A priori}, we cannot drop 
higher mass dimension terms 
when constructing a full effective field theory (EFT) of a moat regime.
Thus
our model in Eq.~\eqref{eff_lag} only caricatures an
EFT about the global minimum of the dispersion relation \footnote{If higher derivative terms dominate with negative coefficients, then even if $Z >0$, it is possible that $E(p)$ about $p=0$ is a local minimum, but that the global minimum of $E(p)$ is at $p\neq0$.},
valid when 
$\langle (\partial_i \vec{\phi})^2\rangle \ll M^4$ and 
$\langle (\vec{\phi})^2\rangle \ll M^2$.
We cannot rule out additional local minima \footnote{This occurs, e.g., for a roton, where $p = 0$
is the global minimum, but there is a local minimum for one or more
$p_{\rm roton} \neq 0$.}; however,
the dominant effects stem from field configurations at the global minimum,
as in Refs. \cite{Pisarski:2020gkx, Pisarski:2021qof,Rennecke:2023xhc}, which our model captures.
For simplicity, we work in the symmetric ($\langle \vec{\phi}\, \rangle = 0$)  regime.\\

\noindent {\bf Dilepton production.} 
At leading order in
$\alpha = e^2/4 \pi$, the dilepton production rate 
is proportional to the photon self energy
$\photonSelfE^{\mu \nu}$ \cite{McLerran:1984ay}:
\begin{equation}\label{eq:dilepton-production}
    \frac{d^4 R}{ dp^4} = \frac{\alpha}{12 \pi^4} \; 
    \frac{n(\omega)}{-P^2}\;  
    \mathrm{Im} \; \photonSelfE^{\mu \mu}(p_0, \vec{p})
    \; .
\end{equation}
Here $\omega > 0$ 
with the Bose function $n(\omega) = 1/({\rm exp}(\omega/T) -1 )$. 
We analytically
continued 
 the Euclidean energy $p_0$ to Minkowski space,  
$p_0 = - i (\omega + i \epsilon)$, where $\epsilon \to 0^+$
so $-P^2 = P_M^2 = \omega^2 - \vec{p}^{\, 2}$, where $P_M$ is a Minkowski $4$-vector.

In vacuum, vector meson dominance implies that pion production
proceeds entirely through virtual $\rho_\mu$ mesons
\cite{OConnell:1995nse}. In heavy ion collisions, successful
fits to the dilepton excess below the
$\rho_\mu$ meson peak are usually obtained by a broadened $\rho_\mu$ in nuclear matter
\cite{Rapp:1999ej,Rapp:2016xzw,Rapp:2024grb}.
In contrast, our analysis is valid in the chirally restored phase, cf.\ Fig.~\ref{fig:sketch}.
Here, we study additional contributions from moatonic pions.

We consider the possible dimension-6 terms that can contribute to dilepton production in a QCD moat regime in four dimensions, where a spin zero field $\vec{\phi}$ carries dimensions of mass. We compare this to the chiral perturbation theory Lagrangian in the Supplementary material.
To obey gauge invariance of couplings to the 
photon field $(A_\mu)$, we replace
$\partial_\mu \rightarrow D_\mu$ in Eq.~\eqref{eff_lag},
\begin{equation}
  \frac{1}{2} (D_0 \vec{\phi})^2  + \frac{Z}{2} (D_i \vec{\phi})^2 \;
  + \frac{1}{2 M^2} (D_i^2 \vec{\phi})^2
  \,,
  \label{der_ints1}
\end{equation}
where the covariant derivative acts as
$D_\mu\vec\phi =(\partial_\mu\sigma,\,\partial_\mu\pi^1-eA_\mu\pi^2,\,\partial_\mu\pi^2+eA_\mu\pi^1,\,\partial_\mu\pi^3)$.
This can be written as 
$D_\mu \cdot \vec{\phi} = (\partial_\mu  - i e t_2 A_\mu
\cdot ) \vec{\phi}$ ;
$t_2$ is the corresponding Pauli matrix, which acts only on 
the $\pi^1$ and $\pi^2$ directions, and is zero 
otherwise; these form charged pions as
 $\pi^\pm = (\pi^1 \pm i \pi^2)/\sqrt{2}$.
\footnote{
This can be derived in a matrix form, where 
$\Phi = \sigma + i \pi^a \sigma^a$, with $\sigma^a$ the Pauli matrices.
Under a gauge transformation,
$\Phi \rightarrow {\rm e}^{i e \theta \sigma^3} 
\Phi {\rm e}^{- i e \theta \sigma^3}$, 
with $D_\mu = \partial_\mu - i e A_\mu [ \sigma^3, ]$. 
For example, charged pions contribute to the usual kinetic term as
$|(\partial_\mu - i e A_\mu) \pi^+|^2$.
}.

There are additional terms involving two derivatives and 
four $\vec{\phi}$'s \cite{Pisarski:2020dnx},
\begin{align}
  & \frac{C_1}{M^2} (D_0 \vec{\phi})^2 \vec{\phi}^{\, 2} + \frac{C_2}{M^2} 
  (D_i \vec{\phi})^2\vec{\phi}^{\, 2}
  \nonumber\\
  &  + \frac{C_3}{M^2} (\vec{\phi} \cdot D_0 \vec{\phi})^{\, 2} 
  + \frac{C_4}{M^2}(\vec{\phi} \cdot D_i\vec{\phi})^2 \, .
  \label{der_ints2}
\end{align}
Since the thermal bath provides a preferred frame of reference,
the coefficients of terms involving temporal and spatial derivatives may differ. Gauge invariance requires that  photons  only  couple through the field strength, $F_{\mu \nu} = \partial_\mu A_\nu - \partial_\nu A_\mu$, giving 
\begin{equation}
  \left(\frac{C_5}{M^2} F_{0 i}^2 + \frac{C_6}{ M^2} F_{i j}^2 \right)
  \vec{\phi}^{\, 2} \; .
\label{der_ints3}
\end{equation}
The above terms are mundane, but there is also
\begin{align}
&\frac{ e\, C_7}{M^2}\, F_{0 i} \left( D_0 \vec{\phi} \cdot
i t_2 \cdot D_{i} \vec{\phi}\right) 
&+ \frac{ e\, C_8}{M^2}\, F_{i j} \left( D_i \vec{\phi} \cdot 
i t_2 \cdot
D_j \vec{\phi} \right) \; .
\label{der_ints4}
\end{align}
In terms of charged pions,
\begin{align}
\frac{i e\, C_7}{M^2} &
F_{0 i}\, \left( D_0 \pi^+ D_i \pi^- - D_i \pi^+ D_0 \pi^-
\right)
\nonumber \\
&+ \frac{i e\, C_8}{M^2} F_{i j} \left(\, D_i \pi^+ D_j \pi^-
\right)\; .
\end{align}
In this form their invariance under charge conjugation is clear, 
as then $A_\mu \rightarrow - A_\mu$ and $\pi^+
\leftrightarrow \pi^-$.
These terms are invariant under time reversal as they have an even number of powers of $\partial_0$ and $A_0$
\footnote{
There is also a term which is odd under charge conjugation,
$\sim F_{0 i} (D_0 \vec{\phi} \cdot D_i \vec{\phi})$.
Unlike the other electromagnetic couplings, this
term couples a photon to both charged and neutral  
(the $\sigma$ and $\pi^3$) fields.  If $\mu \neq 0$ implies net electrical charge,
then the coefficient of this term must vanish at $\mu = 0$.
}.

Eqs.~(\ref{der_ints1})-(\ref{der_ints4}) 
represent the complete set of operators to sixth order
in the mass dimension.  Other terms, 
e.g., 
$(D_i D_j \vec{\phi})\cdot (D_i D_j \vec{\phi})$
and $(D_i D_j \vec{\phi})\cdot (D_j D_i \vec{\phi})$,
are linear combinations of the above.
We emphasize that all these new operators, including the ones in Eq.~\eqref{der_ints1}, are dictated by gauge symmetry. 

%%%%%%%%%%%%%%%%%%%%%
%%%%%%%%%%%%%%%%%%%%%
\begin{figure*}
    \centering
    \includegraphics[trim=0cm 0.4cm 0cm 0.3cm,
    clip,
    keepaspectratio]{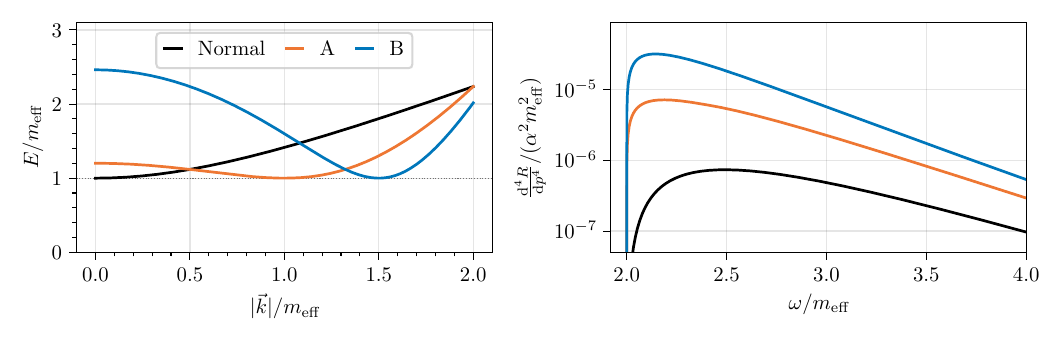}
   \caption{(Left) Dispersion relations for various choices of $Z$ and $M$ in Eq.~\eqref{propagator} and (Right) their corresponding dilepton production rate in Eq.~\eqref{eq:dilepton-production} as a function of $\omega$ at $T= 0.5 m_{\rm eff}$. The normal dispersion relation has $Z=1$, $M = \infty$ and $m_\mathrm{eff}=m$. Both pions with a shallow moat (curve A, with $k_M / m_{\rm eff} = 1.0$ and $M /  m_{\rm eff} = 1.5  $) and pions with a deep moat (curve B, with $k_M / m_{\rm eff} = 1.5$ and $M /  m_{\rm eff} = 1.0  $) {\it dramatically}
   enhances dilepton production.  Note, however, that the value of $m_{\rm eff}$
   {\it does} change as $T$ and $\mu$ vary. 
   As discussed in the text, the terms $C_{1-6}$ never contribute,
   while we set $C_7 = C_8=0$.}
   \label{fig:dispersionAndDilepton}
\end{figure*}
%%%%%%%%%%%%%%%%%%%%%
%%%%%%%%%%%%%%%%%%%%%  

Next, we examine which operators in Eqs. \eqref{der_ints1}-\eqref{der_ints4} contribute to Eq. \eqref{eq:dilepton-production}.
Only the
imaginary part of $\Pi^{\mu\nu}$ enters Eq.~\eqref{eq:dilepton-production}, so
we can drop the terms in Eqs.~\eqref{der_ints2}-\eqref{der_ints3}, as they only affect
tadpole-like diagrams.
There is a contribution to dilepton production
from the $(D_i^2 \vec{\phi})^2$ term in Eq.~\eqref{der_ints1}, 
through the trilinear coupling between
$A_i(P)$, 
$\pi^+(K)$, and $\pi^-(-P-K)$.  This vertex is
\begin{equation}
  \Gamma^j(P,K) = i e\! \left( 2 k + p \right)^j \!\left[Z+   \frac{1}{M^2}\! \Big(\vec{k}^{\, 2} +
  \big(\vec{k} + \vec{p}\big)^2\Big) \right] \, .
  \label{trilinear_vertex}
\end{equation}
From Eq. (\ref{eff_lag}), the (effective) pion propagator in the moat
regime is 
\begin{align}
  \Delta^{-1}(K) &= k_0^2  + Z \vec{k}^{\, 2} +  \frac{(\vec{k}^2)^2}{M^2} + m^2 \label{propagator}
 \\ &= k_0^2 + \frac{\left(\vec{k}^2 - k_M^2\right)^2}{M^2} + m_{\mathrm{eff}}^2 \; ,\nonumber
\end{align}
where the moat momentum $k_M$ satisfies $k_M^2 = - Z M^2 / 2$, and the effective mass is
$m_{\mathrm{eff}}^2 = m^2 - Z^2 M^2 / 4$.
A moat arises when
$Z$ is negative and $k_M^2$ is positive.  

It is readily verified
that the vertex in
Eq. \eqref{trilinear_vertex} satisfies the Ward identity
\begin{equation}
  P^\mu \, \Gamma^\mu(P,K) = i e \left[ \Delta^{-1}(K+P) - \Delta^{-1}(K) \right] \; .
  \label{ward_identity}
\end{equation}
The Ward identity 
implies that the photon self-energy is always transverse,
$P^\mu\,  \photonSelfE^{\mu \nu}(P) = 0$.
Note that this would not be the case without the new contribution $\sim 1/M^2$. 

In this Letter we compute   
dilepton pair production at rest in the frame
of the thermal medium, with $\vec{p} = 0$,
where only spatial components, $\mathrm{Im}\, \photonSelfE^{ii}$, enter.  
To illustrate the effect, we only include the coupling
to photons from Eq. (\ref{der_ints1}), and set the couplings
$C_7 = C_8 = 0$.  The case of
$\vec{p} \neq 0$, with these new couplings,
will be analyzed in future work.

The leading 
contributions to ${\rm Im}\,\, \Pi^{\mu\nu}$ come from 
the moaton-anti-moaton production diagram,
\begin{equation}
    \Pi^{i i}(p_0) = e^2 T \sum_{n'} \int \frac{\dr^3 k}{(2\pi)^3} \left(\Gamma^i\right)^2 \, \Delta(P+K) \, \Delta(P) \; ,
    \label{eq:loop}
\end{equation}
where $K=(2 \pi n' T ,\vec k)$.
Evaluating the bosonic Matsubara frequency sum over $n'$,
\begin{align}
    {\rm Im}\,\Pi_{jj}(\omega) ={}& 4\alpha \theta(\omega^2- 4 m_{\rm eff}^2) \left[1+2 n\!\left(\frac{\omega}{2}\right)\right]
    \nonumber \\
    &\quad \times \frac{  \sqrt{\omega^2 - 4 m_{\rm eff}^2}}{ M \, \omega } \sum_{k_\pm}  k_\pm^3 \theta(k_\pm)
\label{eq:photonselfenergy_moat}
\end{align}
where
\begin{equation}
    k_\pm^2 = k_M^2 \pm 
    \frac{M \sqrt{\omega^2 - 4 m_{\rm eff}^2}  }{2}  \;
\end{equation}
is the spatial momentum of the moatons.  As the moatons
are back to back, their momenta have the same magnitude.
Near threshold, where
$\omega^2 \approx 4 m_{\rm eff}^2- i\epsilon$, we can approximate  
$\vec{k} \approx k_M \widehat{k} + \delta \vec{k}$,
where $\widehat{k}^2 = 1$.
About this point, the energy is quadratic in 
$\delta \vec{k}$,
\begin{equation}
    E_\pm(\vec{k}) \approx m_{\rm eff} +
    \frac{2 \, k_M^2}{m_{\rm eff}\, M^2} (\widehat{k} \cdot \delta \vec{k})^2 
    + \ldots 
    \end{equation}
For pions with a normal dispersion relation
$E^2 = k^2 + m^2$, the photon self-energy is
\begin{align}
    {\rm Im}\,\Pi_{jj}\!=\! \frac{\alpha}{2}  \theta(\omega^2 - 4m^2) \!\left[1+2n\!\left(\frac{\omega}{2}\right)\right]\! %\times 
    %\nonumber\\
    \frac{(\omega^2 - 4m^2)^{\frac{3}{2}}} {\omega}\!.
    \label{eq:normal}
\end{align}

The result for dilepton production from moatons, 
Eq. (\ref{eq:photonselfenergy_moat}), obscures
an interesting cancellation.
For $\vec{p} = 0$, for the imaginary part of the loop integral,
energy momentum conservation imposes
\begin{equation}
    \delta(\omega^2 - 4 m_{\rm eff}^2) = 
    \frac{M^2}{2 k_M^2} \frac{\delta(k - k_\pm)}{|\hat{k} \cdot \delta \vec{k}|} \; ,
    \label{delta_sing}
\end{equation}
where the numerator on the right hand side arises from the Jacobian going from a $\delta$-function in $\omega$ to one in $k$.
This factor would produce a van-Hove singularity, as arises for dilepton production from plasminos in the hard thermal loop approximation \cite{Braaten:1990wp}.
For moatons near threshold, however, 
\begin{equation}
    \Gamma^j(P,K) \approx 2i e k^j
    \left[ \frac{4 k_M}{M^2} ( \hat{k} \cdot \delta \vec{k}) \right]
    \label{vertex_canc}
    \; 
\end{equation}
vanishes as $\delta \vec{k} \rightarrow 0$,
and there is no van Hove singularity.  Consequently
the dilepton rate turns on smoothly at threshold. 
Preliminary numerical computations of Eq.~\eqref{eq:dilepton-production} at $\vec p \neq 0$ indicate that the threshold behavior with contributions from time-like moatons remains qualitatively similar.
We will report on these investigations in a future publication.

The left panel of Fig. (\ref{fig:dispersionAndDilepton}) illustrates three
dispersion relations: normal, a shallow moat,
and a deep moat;
the right panel illustrates their corresponding dilepton production rates. 
These curves have qualitatively similar shapes,
but the dilepton rate is {\it significantly}
enhanced for moatons, regardless of the moat depth. 
The moaton threshold occurs at a nonzero spatial momentum $k_M$ with phase space $\sim k_M^2 dk$,
whereas the normal pion threshold is at $k=0$.

The detailed contribution of moatons to dilepton production is dependent upon effective models.  Nevertheless, surely $m_{\rm eff}$
is of order $m_\pi$.  Thus moaton production should appear
as an excess starting at energies $\sim 2 m_\pi$, above
the usual contributions from a broadened $\rho_\mu$-meson
\cite{Rapp:1999ej,Rapp:2000zd, Rapp:2016xzw}.  This should
be a distinctive experimental signature.
We note that there is also dilepton production near second-order transitions, which generates an excess starting from zero energy ~\cite{Nishimura:2022mku,Nishimura:2023oqn,Nishimura:2024kvz}.
This is kinematically distinct; also, for the reasons discussed above, that from moatons is enhanced over such contributions.\\

\noindent {\bf Outlook.} 
A large body of literature suggests that 
QCD probably exhibits a moat regime in a broad swath
of the $T$-$\mu$ plane.
In this Letter, we showed that enhancement of dilepton production is an important signature of a moat regime.
Our principal shortcoming is that we cannot
predict values for parameters like $M$ and $Z$,
and thus $k_M$ and $m_{\rm eff}$.
Importantly, the value of $m_{\rm eff}$ depends
{\it strongly} upon $T$ and $\mu$, and can only be computed
in an effective model.
We expect that $M$ is a few times $m_\pi$, and $m_{\rm eff}$ is moderately less than $m_\pi$. Work is 
in progress on extracting these parameters using the FRG \cite{frg_moat}.  This shows that the moat regime is in the spacelike regime, with a significant width for the spectral density.  In future work, we will extend our results 
to $p \neq 0$.
It will be interesting to see if a van Hove singularity, which is washed out at $\vec{p} = 0$ by the vertices in Eq. (\ref{vertex_canc}), emerges when $p \neq 0$.
Inevitably, this singularity is smoothed out by the finite width of moatons.

Assuming that freezeout occurs in the moat regime, moat signals from dilepton production and HBT interferometry \cite{Pisarski:2021qof,Rennecke:2023xhc} 
should be observable at the BES at RHIC and CBM at FAIR.   

\acknowledgments
L.P.\ acknowledges discussions with G.\ Markó;
R.D.P., with T. Kunihiro and M. Kitazawa;
S.T.S.,  with I.\ Stewart; 
F.R. and R.D.P., for
stimulating discussions with K. Hayashi, 
and emails with him and Y.\ Tsue.
Z.N.\ was supported by a Leverhulme Trust International
Professorship grant number LIP-202-014 at Oxford (S. L. Sondhi). R.D.P. is supported by the U.S. Department of Energy under contract DE-SC0012704, and thanks the Alexander v.\ Humboldt
Foundation for their support.
L.P., F.R.\ and M.W.\ are supported by the Deutsche Forschungsgemeinschaft (DFG, German Research Foundation) through the Collaborative
Research Center TransRegio CRC-TR 211 “Strong-interaction matter under extreme conditions” – project number
315477589 – TRR 211.
M.W.\ acknowledges the support of the Helmholtz Graduate School for Hadron and Ion Research.
L.P. acknowledges the support of the Giersch Foundation.
S.T.S. was supported by the U.S. Department of Energy, Office of Science, Office of Nuclear Physics from DE-SC0011090; the U.S. National Science Foundation through a Graduate Research Fellowship under Grant No. 1745302; fellowships from the MIT Physics Department and School of Science; and the Hoffman Distinguished Postdoctoral Fellowship through the LDRD Program of Los Alamos National Laboratory under Project 20240786PRD1. Los Alamos National Laboratory is operated by Triad National Security, LLC, for the National Nuclear Security Administration of the U.S. Department of Energy (Contract Nr. 892332188CNA000001).

\appendix

\hfill
\bibliography{combined}

\begin{thebibliography}{122}%
\makeatletter
\providecommand \@ifxundefined [1]{%
 \@ifx{#1\undefined}
}%
\providecommand \@ifnum [1]{%
 \ifnum #1\expandafter \@firstoftwo
 \else \expandafter \@secondoftwo
 \fi
}%
\providecommand \@ifx [1]{%
 \ifx #1\expandafter \@firstoftwo
 \else \expandafter \@secondoftwo
 \fi
}%
\providecommand \natexlab [1]{#1}%
\providecommand \enquote  [1]{``#1''}%
\providecommand \bibnamefont  [1]{#1}%
\providecommand \bibfnamefont [1]{#1}%
\providecommand \citenamefont [1]{#1}%
\providecommand \href@noop [0]{\@secondoftwo}%
\providecommand \href [0]{\begingroup \@sanitize@url \@href}%
\providecommand \@href[1]{\@@startlink{#1}\@@href}%
\providecommand \@@href[1]{\endgroup#1\@@endlink}%
\providecommand \@sanitize@url [0]{\catcode `\\12\catcode `\$12\catcode
  `\&12\catcode `\#12\catcode `\^12\catcode `\_12\catcode `\%12\relax}%
\providecommand \@@startlink[1]{}%
\providecommand \@@endlink[0]{}%
\providecommand \url  [0]{\begingroup\@sanitize@url \@url }%
\providecommand \@url [1]{\endgroup\@href {#1}{\urlprefix }}%
\providecommand \urlprefix  [0]{URL }%
\providecommand \Eprint [0]{\href }%
\providecommand \doibase [0]{https://doi.org/}%
\providecommand \selectlanguage [0]{\@gobble}%
\providecommand \bibinfo  [0]{\@secondoftwo}%
\providecommand \bibfield  [0]{\@secondoftwo}%
\providecommand \translation [1]{[#1]}%
\providecommand \BibitemOpen [0]{}%
\providecommand \bibitemStop [0]{}%
\providecommand \bibitemNoStop [0]{.\EOS\space}%
\providecommand \EOS [0]{\spacefactor3000\relax}%
\providecommand \BibitemShut  [1]{\csname bibitem#1\endcsname}%
\let\auto@bib@innerbib\@empty
%</preamble>
\bibitem [{\citenamefont {Bazavov}\ \emph {et~al.}(2019)\citenamefont {Bazavov}
  \emph {et~al.}}]{HotQCD:2018pds}%
  \BibitemOpen
  \bibfield  {author} {\bibinfo {author} {\bibfnamefont {A.}~\bibnamefont
  {Bazavov}} \emph {et~al.} (\bibinfo {collaboration} {HotQCD}),\ }\bibfield
  {title} {\bibinfo {title} {{Chiral crossover in QCD at zero and non-zero
  chemical potentials}},\ }\href
  {https://doi.org/10.1016/j.physletb.2019.05.013} {\bibfield  {journal}
  {\bibinfo  {journal} {Phys. Lett. B}\ }\textbf {\bibinfo {volume} {795}},\
  \bibinfo {pages} {15} (\bibinfo {year} {2019})},\ \Eprint
  {https://arxiv.org/abs/1812.08235} {arXiv:1812.08235 [hep-lat]} \BibitemShut
  {NoStop}%
\bibitem [{\citenamefont {Borsanyi}\ \emph {et~al.}(2020)\citenamefont
  {Borsanyi}, \citenamefont {Fodor}, \citenamefont {Guenther}, \citenamefont
  {Kara}, \citenamefont {Katz}, \citenamefont {Parotto}, \citenamefont
  {Pasztor}, \citenamefont {Ratti},\ and\ \citenamefont
  {Szabo}}]{Borsanyi:2020fev}%
  \BibitemOpen
  \bibfield  {author} {\bibinfo {author} {\bibfnamefont {S.}~\bibnamefont
  {Borsanyi}}, \bibinfo {author} {\bibfnamefont {Z.}~\bibnamefont {Fodor}},
  \bibinfo {author} {\bibfnamefont {J.~N.}\ \bibnamefont {Guenther}}, \bibinfo
  {author} {\bibfnamefont {R.}~\bibnamefont {Kara}}, \bibinfo {author}
  {\bibfnamefont {S.~D.}\ \bibnamefont {Katz}}, \bibinfo {author}
  {\bibfnamefont {P.~l.}\ \bibnamefont {Parotto}}, \bibinfo {author}
  {\bibfnamefont {A.}~\bibnamefont {Pasztor}}, \bibinfo {author} {\bibfnamefont
  {C.}~\bibnamefont {Ratti}},\ and\ \bibinfo {author} {\bibfnamefont {K.~K.}\
  \bibnamefont {Szabo}},\ }\bibfield  {title} {\bibinfo {title} {{QCD Crossover
  at Finite Chemical Potential from Lattice Simulations}},\ }\href
  {https://doi.org/10.1103/PhysRevLett.125.052001} {\bibfield  {journal}
  {\bibinfo  {journal} {Phys. Rev. Lett.}\ }\textbf {\bibinfo {volume} {125}},\
  \bibinfo {pages} {052001} (\bibinfo {year} {2020})},\ \Eprint
  {https://arxiv.org/abs/2002.02821} {arXiv:2002.02821 [hep-lat]} \BibitemShut
  {NoStop}%
\bibitem [{\citenamefont {Asakawa}\ and\ \citenamefont
  {Yazaki}(1989)}]{Asakawa:1989bq}%
  \BibitemOpen
  \bibfield  {author} {\bibinfo {author} {\bibfnamefont {M.}~\bibnamefont
  {Asakawa}}\ and\ \bibinfo {author} {\bibfnamefont {K.}~\bibnamefont
  {Yazaki}},\ }\bibfield  {title} {\bibinfo {title} {{Chiral Restoration at
  Finite Density and Temperature}},\ }\href
  {https://doi.org/10.1016/0375-9474(89)90002-X} {\bibfield  {journal}
  {\bibinfo  {journal} {Nucl. Phys.}\ }\textbf {\bibinfo {volume} {A504}},\
  \bibinfo {pages} {668} (\bibinfo {year} {1989})}\BibitemShut {NoStop}%
%%CITATION = NUPHA,A504,668;%%
\bibitem [{\citenamefont {Stephanov}\ \emph {et~al.}(1998)\citenamefont
  {Stephanov}, \citenamefont {Rajagopal},\ and\ \citenamefont
  {Shuryak}}]{Stephanov:1998dy}%
  \BibitemOpen
  \bibfield  {author} {\bibinfo {author} {\bibfnamefont {M.~A.}\ \bibnamefont
  {Stephanov}}, \bibinfo {author} {\bibfnamefont {K.}~\bibnamefont
  {Rajagopal}},\ and\ \bibinfo {author} {\bibfnamefont {E.~V.}\ \bibnamefont
  {Shuryak}},\ }\bibfield  {title} {\bibinfo {title} {{Signatures of the
  tricritical point in QCD}},\ }\href
  {https://doi.org/10.1103/PhysRevLett.81.4816} {\bibfield  {journal} {\bibinfo
   {journal} {Phys. Rev. Lett.}\ }\textbf {\bibinfo {volume} {81}},\ \bibinfo
  {pages} {4816} (\bibinfo {year} {1998})},\ \Eprint
  {https://arxiv.org/abs/hep-ph/9806219} {arXiv:hep-ph/9806219 [hep-ph]}
  \BibitemShut {NoStop}%
%%CITATION = HEP-PH/9806219;%%
\bibitem [{\citenamefont {Stephanov}\ \emph {et~al.}(1999)\citenamefont
  {Stephanov}, \citenamefont {Rajagopal},\ and\ \citenamefont
  {Shuryak}}]{Stephanov:1999zu}%
  \BibitemOpen
  \bibfield  {author} {\bibinfo {author} {\bibfnamefont {M.~A.}\ \bibnamefont
  {Stephanov}}, \bibinfo {author} {\bibfnamefont {K.}~\bibnamefont
  {Rajagopal}},\ and\ \bibinfo {author} {\bibfnamefont {E.~V.}\ \bibnamefont
  {Shuryak}},\ }\bibfield  {title} {\bibinfo {title} {{Event-by-event
  fluctuations in heavy ion collisions and the QCD critical point}},\ }\href
  {https://doi.org/10.1103/PhysRevD.60.114028} {\bibfield  {journal} {\bibinfo
  {journal} {Phys. Rev.}\ }\textbf {\bibinfo {volume} {D60}},\ \bibinfo {pages}
  {114028} (\bibinfo {year} {1999})},\ \Eprint
  {https://arxiv.org/abs/hep-ph/9903292} {arXiv:hep-ph/9903292 [hep-ph]}
  \BibitemShut {NoStop}%
%%CITATION = HEP-PH/9903292;%%
\bibitem [{\citenamefont {Parotto}\ \emph {et~al.}(2020)\citenamefont
  {Parotto}, \citenamefont {Bluhm}, \citenamefont {Mroczek}, \citenamefont
  {Nahrgang}, \citenamefont {Noronha-Hostler}, \citenamefont {Rajagopal},
  \citenamefont {Ratti}, \citenamefont {Sch\"afer},\ and\ \citenamefont
  {Stephanov}}]{Parotto:2018pwx}%
  \BibitemOpen
  \bibfield  {author} {\bibinfo {author} {\bibfnamefont {P.}~\bibnamefont
  {Parotto}}, \bibinfo {author} {\bibfnamefont {M.}~\bibnamefont {Bluhm}},
  \bibinfo {author} {\bibfnamefont {D.}~\bibnamefont {Mroczek}}, \bibinfo
  {author} {\bibfnamefont {M.}~\bibnamefont {Nahrgang}}, \bibinfo {author}
  {\bibfnamefont {J.}~\bibnamefont {Noronha-Hostler}}, \bibinfo {author}
  {\bibfnamefont {K.}~\bibnamefont {Rajagopal}}, \bibinfo {author}
  {\bibfnamefont {C.}~\bibnamefont {Ratti}}, \bibinfo {author} {\bibfnamefont
  {T.}~\bibnamefont {Sch\"afer}},\ and\ \bibinfo {author} {\bibfnamefont
  {M.}~\bibnamefont {Stephanov}},\ }\bibfield  {title} {\bibinfo {title} {{QCD
  equation of state matched to lattice data and exhibiting a critical point
  singularity}},\ }\href {https://doi.org/10.1103/PhysRevC.101.034901}
  {\bibfield  {journal} {\bibinfo  {journal} {Phys. Rev. C}\ }\textbf {\bibinfo
  {volume} {101}},\ \bibinfo {pages} {034901} (\bibinfo {year} {2020})},\
  \Eprint {https://arxiv.org/abs/1805.05249} {arXiv:1805.05249 [hep-ph]}
  \BibitemShut {NoStop}%
\bibitem [{\citenamefont {Bzdak}\ \emph {et~al.}(2020)\citenamefont {Bzdak},
  \citenamefont {Esumi}, \citenamefont {Koch}, \citenamefont {Liao},
  \citenamefont {Stephanov},\ and\ \citenamefont {Xu}}]{Bzdak:2019pkr}%
  \BibitemOpen
  \bibfield  {author} {\bibinfo {author} {\bibfnamefont {A.}~\bibnamefont
  {Bzdak}}, \bibinfo {author} {\bibfnamefont {S.}~\bibnamefont {Esumi}},
  \bibinfo {author} {\bibfnamefont {V.}~\bibnamefont {Koch}}, \bibinfo {author}
  {\bibfnamefont {J.}~\bibnamefont {Liao}}, \bibinfo {author} {\bibfnamefont
  {M.}~\bibnamefont {Stephanov}},\ and\ \bibinfo {author} {\bibfnamefont
  {N.}~\bibnamefont {Xu}},\ }\bibfield  {title} {\bibinfo {title} {{Mapping the
  Phases of Quantum Chromodynamics with Beam Energy Scan}},\ }\href
  {https://doi.org/10.1016/j.physrep.2020.01.005} {\bibfield  {journal}
  {\bibinfo  {journal} {Phys. Rept.}\ }\textbf {\bibinfo {volume} {853}},\
  \bibinfo {pages} {1} (\bibinfo {year} {2020})},\ \Eprint
  {https://arxiv.org/abs/1906.00936} {arXiv:1906.00936 [nucl-th]} \BibitemShut
  {NoStop}%
\bibitem [{\citenamefont {Fu}\ \emph {et~al.}(2020)\citenamefont {Fu},
  \citenamefont {Pawlowski},\ and\ \citenamefont {Rennecke}}]{Fu:2019hdw}%
  \BibitemOpen
  \bibfield  {author} {\bibinfo {author} {\bibfnamefont {W.-j.}\ \bibnamefont
  {Fu}}, \bibinfo {author} {\bibfnamefont {J.~M.}\ \bibnamefont {Pawlowski}},\
  and\ \bibinfo {author} {\bibfnamefont {F.}~\bibnamefont {Rennecke}},\
  }\bibfield  {title} {\bibinfo {title} {The {{QCD}} phase structure at finite
  temperature and density},\ }\href
  {https://doi.org/10.1103/PhysRevD.101.054032} {\bibfield  {journal} {\bibinfo
   {journal} {Physical Review D}\ }\textbf {\bibinfo {volume} {101}},\ \bibinfo
  {pages} {054032} (\bibinfo {year} {2020})},\ \Eprint
  {https://arxiv.org/abs/1909.02991} {arxiv:1909.02991} \BibitemShut {NoStop}%
\bibitem [{\citenamefont {Gao}\ and\ \citenamefont
  {Pawlowski}(2020)}]{Gao:2020qsj}%
  \BibitemOpen
  \bibfield  {author} {\bibinfo {author} {\bibfnamefont {F.}~\bibnamefont
  {Gao}}\ and\ \bibinfo {author} {\bibfnamefont {J.~M.}\ \bibnamefont
  {Pawlowski}},\ }\bibfield  {title} {\bibinfo {title} {{QCD phase structure
  from functional methods}},\ }\href
  {https://doi.org/10.1103/PhysRevD.102.034027} {\bibfield  {journal} {\bibinfo
   {journal} {Phys. Rev. D}\ }\textbf {\bibinfo {volume} {102}},\ \bibinfo
  {pages} {034027} (\bibinfo {year} {2020})},\ \Eprint
  {https://arxiv.org/abs/2002.07500} {arXiv:2002.07500 [hep-ph]} \BibitemShut
  {NoStop}%
\bibitem [{\citenamefont {Gao}\ and\ \citenamefont
  {Pawlowski}(2021)}]{Gao:2020fbl}%
  \BibitemOpen
  \bibfield  {author} {\bibinfo {author} {\bibfnamefont {F.}~\bibnamefont
  {Gao}}\ and\ \bibinfo {author} {\bibfnamefont {J.~M.}\ \bibnamefont
  {Pawlowski}},\ }\bibfield  {title} {\bibinfo {title} {Chiral phase structure
  and critical end point in {{QCD}}},\ }\href
  {https://doi.org/10.1016/j.physletb.2021.136584} {\bibfield  {journal}
  {\bibinfo  {journal} {Physics Letters B}\ }\textbf {\bibinfo {volume}
  {820}},\ \bibinfo {pages} {136584} (\bibinfo {year} {2021})},\ \Eprint
  {https://arxiv.org/abs/2010.13705} {2010.13705 [hep-ph, physics:hep-th]}
  \BibitemShut {NoStop}%
\bibitem [{\citenamefont {Gunkel}\ and\ \citenamefont
  {Fischer}(2021)}]{Gunkel:2021oya}%
  \BibitemOpen
  \bibfield  {author} {\bibinfo {author} {\bibfnamefont {P.~J.}\ \bibnamefont
  {Gunkel}}\ and\ \bibinfo {author} {\bibfnamefont {C.~S.}\ \bibnamefont
  {Fischer}},\ }\bibfield  {title} {\bibinfo {title} {{Locating the critical
  endpoint of QCD: Mesonic backcoupling effects}},\ }\href
  {https://doi.org/10.1103/PhysRevD.104.054022} {\bibfield  {journal} {\bibinfo
   {journal} {Phys. Rev. D}\ }\textbf {\bibinfo {volume} {104}},\ \bibinfo
  {pages} {054022} (\bibinfo {year} {2021})},\ \Eprint
  {https://arxiv.org/abs/2106.08356} {arXiv:2106.08356 [hep-ph]} \BibitemShut
  {NoStop}%
\bibitem [{\citenamefont {Basar}(2024)}]{Basar:2023nkp}%
  \BibitemOpen
  \bibfield  {author} {\bibinfo {author} {\bibfnamefont {G.}~\bibnamefont
  {Basar}},\ }\bibfield  {title} {\bibinfo {title} {{QCD critical point,
  Lee-Yang edge singularities, and Pad\'e resummations}},\ }\href
  {https://doi.org/10.1103/PhysRevC.110.015203} {\bibfield  {journal} {\bibinfo
   {journal} {Phys. Rev. C}\ }\textbf {\bibinfo {volume} {110}},\ \bibinfo
  {pages} {015203} (\bibinfo {year} {2024})},\ \Eprint
  {https://arxiv.org/abs/2312.06952} {arXiv:2312.06952 [hep-th]} \BibitemShut
  {NoStop}%
\bibitem [{\citenamefont {Clarke}\ \emph {et~al.}(2024)\citenamefont {Clarke},
  \citenamefont {Dimopoulos}, \citenamefont {Di~Renzo}, \citenamefont
  {Goswami}, \citenamefont {Schmidt}, \citenamefont {Singh},\ and\
  \citenamefont {Zambello}}]{Clarke:2024ugt}%
  \BibitemOpen
  \bibfield  {author} {\bibinfo {author} {\bibfnamefont {D.~A.}\ \bibnamefont
  {Clarke}}, \bibinfo {author} {\bibfnamefont {P.}~\bibnamefont {Dimopoulos}},
  \bibinfo {author} {\bibfnamefont {F.}~\bibnamefont {Di~Renzo}}, \bibinfo
  {author} {\bibfnamefont {J.}~\bibnamefont {Goswami}}, \bibinfo {author}
  {\bibfnamefont {C.}~\bibnamefont {Schmidt}}, \bibinfo {author} {\bibfnamefont
  {S.}~\bibnamefont {Singh}},\ and\ \bibinfo {author} {\bibfnamefont
  {K.}~\bibnamefont {Zambello}},\ }\bibfield  {title} {\bibinfo {title}
  {{Searching for the QCD critical endpoint using multi-point Pad\'e
  approximations}},\ }\href@noop {} {\  (\bibinfo {year} {2024})},\ \Eprint
  {https://arxiv.org/abs/2405.10196} {arXiv:2405.10196 [hep-lat]} \BibitemShut
  {NoStop}%
\bibitem [{Note1()}]{Note1}%
  \BibitemOpen
  \bibinfo {note} {To be more precise, the critical mode is a mixture between
  the $\sigma $, the $\omega ^0$ meson and the Polyakov loops, and the relevant
  mass that vanishes at the CEP is the spacelike screening mass \cite
  {Haensch:2023sig}.}\BibitemShut {Stop}%
\bibitem [{\citenamefont {Braun}\ \emph {et~al.}(2011)\citenamefont {Braun},
  \citenamefont {Klein},\ and\ \citenamefont {Piasecki}}]{Braun:2010vd}%
  \BibitemOpen
  \bibfield  {author} {\bibinfo {author} {\bibfnamefont {J.}~\bibnamefont
  {Braun}}, \bibinfo {author} {\bibfnamefont {B.}~\bibnamefont {Klein}},\ and\
  \bibinfo {author} {\bibfnamefont {P.}~\bibnamefont {Piasecki}},\ }\bibfield
  {title} {\bibinfo {title} {{On the scaling behavior of the chiral phase
  transition in QCD in finite and infinite volume}},\ }\href
  {https://doi.org/10.1140/epjc/s10052-011-1576-7} {\bibfield  {journal}
  {\bibinfo  {journal} {Eur. Phys. J. C}\ }\textbf {\bibinfo {volume} {71}},\
  \bibinfo {pages} {1576} (\bibinfo {year} {2011})},\ \Eprint
  {https://arxiv.org/abs/1008.2155} {arXiv:1008.2155 [hep-ph]} \BibitemShut
  {NoStop}%
\bibitem [{\citenamefont {Schaefer}\ and\ \citenamefont
  {Wagner}(2012)}]{Schaefer:2011ex}%
  \BibitemOpen
  \bibfield  {author} {\bibinfo {author} {\bibfnamefont {B.~J.}\ \bibnamefont
  {Schaefer}}\ and\ \bibinfo {author} {\bibfnamefont {M.}~\bibnamefont
  {Wagner}},\ }\bibfield  {title} {\bibinfo {title} {{QCD critical region and
  higher moments for three flavor models}},\ }\href
  {https://doi.org/10.1103/PhysRevD.85.034027} {\bibfield  {journal} {\bibinfo
  {journal} {Phys. Rev. D}\ }\textbf {\bibinfo {volume} {85}},\ \bibinfo
  {pages} {034027} (\bibinfo {year} {2012})},\ \Eprint
  {https://arxiv.org/abs/1111.6871} {arXiv:1111.6871 [hep-ph]} \BibitemShut
  {NoStop}%
\bibitem [{\citenamefont {Braun}\ \emph {et~al.}(2020)\citenamefont {Braun},
  \citenamefont {Fu}, \citenamefont {Pawlowski}, \citenamefont {Rennecke},
  \citenamefont {Rosenbl\"uh},\ and\ \citenamefont {Yin}}]{Braun:2020ada}%
  \BibitemOpen
  \bibfield  {author} {\bibinfo {author} {\bibfnamefont {J.}~\bibnamefont
  {Braun}}, \bibinfo {author} {\bibfnamefont {W.-j.}\ \bibnamefont {Fu}},
  \bibinfo {author} {\bibfnamefont {J.~M.}\ \bibnamefont {Pawlowski}}, \bibinfo
  {author} {\bibfnamefont {F.}~\bibnamefont {Rennecke}}, \bibinfo {author}
  {\bibfnamefont {D.}~\bibnamefont {Rosenbl\"uh}},\ and\ \bibinfo {author}
  {\bibfnamefont {S.}~\bibnamefont {Yin}},\ }\bibfield  {title} {\bibinfo
  {title} {{Chiral susceptibility in ( 2+1 )-flavor QCD}},\ }\href
  {https://doi.org/10.1103/PhysRevD.102.056010} {\bibfield  {journal} {\bibinfo
   {journal} {Phys. Rev. D}\ }\textbf {\bibinfo {volume} {102}},\ \bibinfo
  {pages} {056010} (\bibinfo {year} {2020})},\ \Eprint
  {https://arxiv.org/abs/2003.13112} {arXiv:2003.13112 [hep-ph]} \BibitemShut
  {NoStop}%
\bibitem [{\citenamefont {Gao}\ and\ \citenamefont
  {Pawlowski}(2022)}]{Gao:2021vsf}%
  \BibitemOpen
  \bibfield  {author} {\bibinfo {author} {\bibfnamefont {F.}~\bibnamefont
  {Gao}}\ and\ \bibinfo {author} {\bibfnamefont {J.~M.}\ \bibnamefont
  {Pawlowski}},\ }\bibfield  {title} {\bibinfo {title} {{Phase structure of
  (2+1)-flavor QCD and the magnetic equation of state}},\ }\href
  {https://doi.org/10.1103/PhysRevD.105.094020} {\bibfield  {journal} {\bibinfo
   {journal} {Phys. Rev. D}\ }\textbf {\bibinfo {volume} {105}},\ \bibinfo
  {pages} {094020} (\bibinfo {year} {2022})},\ \Eprint
  {https://arxiv.org/abs/2112.01395} {arXiv:2112.01395 [hep-ph]} \BibitemShut
  {NoStop}%
\bibitem [{\citenamefont {Fu}\ \emph {et~al.}(2023)\citenamefont {Fu},
  \citenamefont {Luo}, \citenamefont {Pawlowski}, \citenamefont {Rennecke},\
  and\ \citenamefont {Yin}}]{Fu:2023lcm}%
  \BibitemOpen
  \bibfield  {author} {\bibinfo {author} {\bibfnamefont {W.-j.}\ \bibnamefont
  {Fu}}, \bibinfo {author} {\bibfnamefont {X.}~\bibnamefont {Luo}}, \bibinfo
  {author} {\bibfnamefont {J.~M.}\ \bibnamefont {Pawlowski}}, \bibinfo {author}
  {\bibfnamefont {F.}~\bibnamefont {Rennecke}},\ and\ \bibinfo {author}
  {\bibfnamefont {S.}~\bibnamefont {Yin}},\ }\bibfield  {title} {\bibinfo
  {title} {{Ripples of the QCD Critical Point}},\ }\href@noop {} {\  (\bibinfo
  {year} {2023})},\ \Eprint {https://arxiv.org/abs/2308.15508}
  {arXiv:2308.15508 [hep-ph]} \BibitemShut {NoStop}%
\bibitem [{\citenamefont {Bernhardt}\ and\ \citenamefont
  {Fischer}(2023)}]{Bernhardt:2023hpr}%
  \BibitemOpen
  \bibfield  {author} {\bibinfo {author} {\bibfnamefont {J.}~\bibnamefont
  {Bernhardt}}\ and\ \bibinfo {author} {\bibfnamefont {C.~S.}\ \bibnamefont
  {Fischer}},\ }\bibfield  {title} {\bibinfo {title} {{QCD phase transitions in
  the light quark chiral limit}},\ }\href
  {https://doi.org/10.1103/PhysRevD.108.114018} {\bibfield  {journal} {\bibinfo
   {journal} {Phys. Rev. D}\ }\textbf {\bibinfo {volume} {108}},\ \bibinfo
  {pages} {114018} (\bibinfo {year} {2023})},\ \Eprint
  {https://arxiv.org/abs/2309.06737} {arXiv:2309.06737 [hep-ph]} \BibitemShut
  {NoStop}%
\bibitem [{\citenamefont {Braun}\ \emph {et~al.}(2023)\citenamefont {Braun}
  \emph {et~al.}}]{Braun:2023qak}%
  \BibitemOpen
  \bibfield  {author} {\bibinfo {author} {\bibfnamefont {J.}~\bibnamefont
  {Braun}} \emph {et~al.},\ }\bibfield  {title} {\bibinfo {title} {{Soft modes
  in hot QCD matter}},\ }\href@noop {} {\  (\bibinfo {year} {2023})},\ \Eprint
  {https://arxiv.org/abs/2310.19853} {arXiv:2310.19853 [hep-ph]} \BibitemShut
  {NoStop}%
\bibitem [{\citenamefont {Lu}\ \emph {et~al.}(2024)\citenamefont {Lu},
  \citenamefont {Gao}, \citenamefont {Liu},\ and\ \citenamefont
  {Pawlowski}}]{Lu:2023mkn}%
  \BibitemOpen
  \bibfield  {author} {\bibinfo {author} {\bibfnamefont {Y.}~\bibnamefont
  {Lu}}, \bibinfo {author} {\bibfnamefont {F.}~\bibnamefont {Gao}}, \bibinfo
  {author} {\bibfnamefont {Y.-X.}\ \bibnamefont {Liu}},\ and\ \bibinfo {author}
  {\bibfnamefont {J.~M.}\ \bibnamefont {Pawlowski}},\ }\bibfield  {title}
  {\bibinfo {title} {{QCD equation of state and thermodynamic observables from
  computationally minimal Dyson-Schwinger equations}},\ }\href
  {https://doi.org/10.1103/PhysRevD.110.014036} {\bibfield  {journal} {\bibinfo
   {journal} {Phys. Rev. D}\ }\textbf {\bibinfo {volume} {110}},\ \bibinfo
  {pages} {014036} (\bibinfo {year} {2024})},\ \Eprint
  {https://arxiv.org/abs/2310.18383} {arXiv:2310.18383 [hep-ph]} \BibitemShut
  {NoStop}%
\bibitem [{\citenamefont {Fu}\ \emph {et~al.}(2024{\natexlab{a}})\citenamefont
  {Fu}, \citenamefont {Pawlowski}, \citenamefont {Pisarski}, \citenamefont
  {Rennecke}, \citenamefont {Wen},\ and\ \citenamefont {Yin}}]{Fu:2024rto}%
  \BibitemOpen
  \bibfield  {author} {\bibinfo {author} {\bibfnamefont {W.-j.}\ \bibnamefont
  {Fu}}, \bibinfo {author} {\bibfnamefont {J.~M.}\ \bibnamefont {Pawlowski}},
  \bibinfo {author} {\bibfnamefont {R.~D.}\ \bibnamefont {Pisarski}}, \bibinfo
  {author} {\bibfnamefont {F.}~\bibnamefont {Rennecke}}, \bibinfo {author}
  {\bibfnamefont {R.}~\bibnamefont {Wen}},\ and\ \bibinfo {author}
  {\bibfnamefont {S.}~\bibnamefont {Yin}},\ }\bibfield  {title} {\bibinfo
  {title} {{The QCD moat regime and its real-time properties}},\ }\href@noop {}
  {\  (\bibinfo {year} {2024}{\natexlab{a}})},\ \Eprint
  {https://arxiv.org/abs/2412.15949} {arXiv:2412.15949 [hep-ph]} \BibitemShut
  {NoStop}%
\bibitem [{\citenamefont {Stephenson}(1970)}]{stephenson_ising_1970}%
  \BibitemOpen
  \bibfield  {author} {\bibinfo {author} {\bibfnamefont {J.}~\bibnamefont
  {Stephenson}},\ }\bibfield  {title} {\bibinfo {title} {Ising {Model} with
  {Antiferromagnetic} {Next}-{Nearest}-{Neighbor} {Coupling}: {Spin}
  {Correlations} and {Disorder} {Points}},\ }\href
  {https://doi.org/10.1103/PhysRevB.1.4405} {\bibfield  {journal} {\bibinfo
  {journal} {Physical Review B}\ }\textbf {\bibinfo {volume} {1}},\ \bibinfo
  {pages} {4405} (\bibinfo {year} {1970})}\BibitemShut {NoStop}%
\bibitem [{\citenamefont {Schindler}\ \emph {et~al.}(2020)\citenamefont
  {Schindler}, \citenamefont {Schindler}, \citenamefont {Medina},\ and\
  \citenamefont {Ogilvie}}]{Schindler:2019ugo}%
  \BibitemOpen
  \bibfield  {author} {\bibinfo {author} {\bibfnamefont {M.~A.}\ \bibnamefont
  {Schindler}}, \bibinfo {author} {\bibfnamefont {S.~T.}\ \bibnamefont
  {Schindler}}, \bibinfo {author} {\bibfnamefont {L.}~\bibnamefont {Medina}},\
  and\ \bibinfo {author} {\bibfnamefont {M.~C.}\ \bibnamefont {Ogilvie}},\
  }\bibfield  {title} {\bibinfo {title} {{Universality of Pattern Formation}},\
  }\href {https://doi.org/10.1103/PhysRevD.102.114510} {\bibfield  {journal}
  {\bibinfo  {journal} {Phys. Rev. D}\ }\textbf {\bibinfo {volume} {102}},\
  \bibinfo {pages} {114510} (\bibinfo {year} {2020})},\ \Eprint
  {https://arxiv.org/abs/1906.07288} {arXiv:1906.07288 [hep-lat]} \BibitemShut
  {NoStop}%
\bibitem [{\citenamefont {Schindler}\ \emph {et~al.}(2021)\citenamefont
  {Schindler}, \citenamefont {Schindler},\ and\ \citenamefont
  {Ogilvie}}]{Schindler:2021otf}%
  \BibitemOpen
  \bibfield  {author} {\bibinfo {author} {\bibfnamefont {M.~A.}\ \bibnamefont
  {Schindler}}, \bibinfo {author} {\bibfnamefont {S.~T.}\ \bibnamefont
  {Schindler}},\ and\ \bibinfo {author} {\bibfnamefont {M.~C.}\ \bibnamefont
  {Ogilvie}},\ }\bibfield  {title} {\bibinfo {title} {{$\mathcal PT$ symmetry,
  pattern formation, and finite-density QCD}},\ }\bibfield  {journal} {\bibinfo
   {journal} {J. Phys.: Conf. Ser.}\ }\textbf {\bibinfo {volume} {2038}},\
  \href {https://doi.org/10.1088/1742-6596/2038/1/012022}
  {10.1088/1742-6596/2038/1/012022} (\bibinfo {year} {2021}),\ \Eprint
  {https://arxiv.org/abs/2106.07092} {arXiv:2106.07092 [hep-lat]} \BibitemShut
  {NoStop}%
\bibitem [{\citenamefont {Fetter}\ and\ \citenamefont
  {Walecka}(2012)}]{fetter2012quantum}%
  \BibitemOpen
  \bibfield  {author} {\bibinfo {author} {\bibfnamefont {A.~L.}\ \bibnamefont
  {Fetter}}\ and\ \bibinfo {author} {\bibfnamefont {J.~D.}\ \bibnamefont
  {Walecka}},\ }\href@noop {} {\emph {\bibinfo {title} {Quantum theory of
  many-particle systems}}}\ (\bibinfo  {publisher} {Courier Corporation},\
  \bibinfo {year} {2012})\BibitemShut {NoStop}%
\bibitem [{\citenamefont {Overhauser}(1960)}]{Overhauser:1960}%
  \BibitemOpen
  \bibfield  {author} {\bibinfo {author} {\bibfnamefont {A.~W.}\ \bibnamefont
  {Overhauser}},\ }\bibfield  {title} {\bibinfo {title} {{Structure of nuclear
  matter}},\ }\href {https://doi.org/10.1103/PhysRevLett.4.415} {\bibfield
  {journal} {\bibinfo  {journal} {Phys. Rev. Lett.}\ }\textbf {\bibinfo
  {volume} {4}},\ \bibinfo {pages} {415} (\bibinfo {year} {1960})}\BibitemShut
  {NoStop}%
\bibitem [{\citenamefont {Migdal}(1971)}]{Migdal:1971}%
  \BibitemOpen
  \bibfield  {author} {\bibinfo {author} {\bibfnamefont {A.~B.}\ \bibnamefont
  {Migdal}},\ }\bibfield  {title} {\bibinfo {title} {{Pion condensation}},\
  }\href@noop {} {\bibfield  {journal} {\bibinfo  {journal} {Zh. Eksp. Teor.
  Fiz}\ }\textbf {\bibinfo {volume} {61}},\ \bibinfo {pages} {2210} (\bibinfo
  {year} {1971})},\ \bibinfo {note} {[Sov. Phys. JETP 36, 1052
  (1973)]}\BibitemShut {NoStop}%
\bibitem [{\citenamefont {Sawyer}(1972)}]{Sawyer:1972cq}%
  \BibitemOpen
  \bibfield  {author} {\bibinfo {author} {\bibfnamefont {R.~F.}\ \bibnamefont
  {Sawyer}},\ }\bibfield  {title} {\bibinfo {title} {{Condensed pion phase in
  neutron star matter}},\ }\href {https://doi.org/10.1103/PhysRevLett.29.382}
  {\bibfield  {journal} {\bibinfo  {journal} {Phys. Rev. Lett.}\ }\textbf
  {\bibinfo {volume} {29}},\ \bibinfo {pages} {382} (\bibinfo {year}
  {1972})}\BibitemShut {NoStop}%
%%CITATION = PRLTA,29,382;%%
\bibitem [{\citenamefont {Migdal}(1973)}]{Migdal:1973zm}%
  \BibitemOpen
  \bibfield  {author} {\bibinfo {author} {\bibfnamefont {A.~B.}\ \bibnamefont
  {Migdal}},\ }\bibfield  {title} {\bibinfo {title} {{Pi condensation in
  nuclear matter}},\ }\href {https://doi.org/10.1103/PhysRevLett.31.257}
  {\bibfield  {journal} {\bibinfo  {journal} {Phys. Rev. Lett.}\ }\textbf
  {\bibinfo {volume} {31}},\ \bibinfo {pages} {257} (\bibinfo {year}
  {1973})}\BibitemShut {NoStop}%
%%CITATION = PRLTA,31,257;%%
\bibitem [{\citenamefont {Kaplan}\ and\ \citenamefont
  {Nelson}(1986)}]{Kaplan:1986yq}%
  \BibitemOpen
  \bibfield  {author} {\bibinfo {author} {\bibfnamefont {D.~B.}\ \bibnamefont
  {Kaplan}}\ and\ \bibinfo {author} {\bibfnamefont {A.~E.}\ \bibnamefont
  {Nelson}},\ }\bibfield  {title} {\bibinfo {title} {{Strange Goings on in
  Dense Nucleonic Matter}},\ }\href
  {https://doi.org/10.1016/0370-2693(86)90331-X} {\bibfield  {journal}
  {\bibinfo  {journal} {Phys. Lett.}\ }\textbf {\bibinfo {volume} {B175}},\
  \bibinfo {pages} {57} (\bibinfo {year} {1986})}\BibitemShut {NoStop}%
%%CITATION = PHLTA,B175,57;%%
\bibitem [{\citenamefont {Kryjevski}(2008)}]{Kryjevski:2005qq}%
  \BibitemOpen
  \bibfield  {author} {\bibinfo {author} {\bibfnamefont {A.}~\bibnamefont
  {Kryjevski}},\ }\bibfield  {title} {\bibinfo {title} {{Spontaneous superfluid
  current generation in the kaon condensed color flavor locked phase at nonzero
  strange quark mass}},\ }\href {https://doi.org/10.1103/PhysRevD.77.014018}
  {\bibfield  {journal} {\bibinfo  {journal} {Phys. Rev. D}\ }\textbf {\bibinfo
  {volume} {77}},\ \bibinfo {pages} {014018} (\bibinfo {year} {2008})},\
  \Eprint {https://arxiv.org/abs/hep-ph/0508180} {arXiv:hep-ph/0508180}
  \BibitemShut {NoStop}%
\bibitem [{\citenamefont {Schaefer}\ and\ \citenamefont
  {Wambach}(2007)}]{Schaefer:2006ds}%
  \BibitemOpen
  \bibfield  {author} {\bibinfo {author} {\bibfnamefont {B.-J.}\ \bibnamefont
  {Schaefer}}\ and\ \bibinfo {author} {\bibfnamefont {J.}~\bibnamefont
  {Wambach}},\ }\bibfield  {title} {\bibinfo {title} {{Susceptibilities near
  the QCD (tri)critical point}},\ }\href
  {https://doi.org/10.1103/PhysRevD.75.085015} {\bibfield  {journal} {\bibinfo
  {journal} {Phys. Rev. D}\ }\textbf {\bibinfo {volume} {75}},\ \bibinfo
  {pages} {085015} (\bibinfo {year} {2007})},\ \Eprint
  {https://arxiv.org/abs/hep-ph/0603256} {arXiv:hep-ph/0603256} \BibitemShut
  {NoStop}%
\bibitem [{\citenamefont {Yoshimori}(1959)}]{Yoshimori}%
  \BibitemOpen
  \bibfield  {author} {\bibinfo {author} {\bibfnamefont {A.}~\bibnamefont
  {Yoshimori}},\ }\bibfield  {title} {\bibinfo {title} {A new type of
  antiferromagnetic structure},\ }\href@noop {} {\bibfield  {journal} {\bibinfo
   {journal} {J. Phys. Soc. Jpn.}\ }\textbf {\bibinfo {volume} {14}},\ \bibinfo
  {pages} {805} (\bibinfo {year} {1959})}\BibitemShut {NoStop}%
\bibitem [{\citenamefont {Hornreich}\ \emph {et~al.}(1975)\citenamefont
  {Hornreich}, \citenamefont {Luban},\ and\ \citenamefont
  {Shtrikman}}]{Hornnerich}%
  \BibitemOpen
  \bibfield  {author} {\bibinfo {author} {\bibfnamefont {R.~M.}\ \bibnamefont
  {Hornreich}}, \bibinfo {author} {\bibfnamefont {M.}~\bibnamefont {Luban}},\
  and\ \bibinfo {author} {\bibfnamefont {S.}~\bibnamefont {Shtrikman}},\
  }\bibfield  {title} {\bibinfo {title} {Critical behavior at the onset of
  $\vec{k}$-space instability on the $\lambda$ line},\ }\href@noop {}
  {\bibfield  {journal} {\bibinfo  {journal} {Phys. Rev. Lett.}\ }\textbf
  {\bibinfo {volume} {35}},\ \bibinfo {pages} {1678} (\bibinfo {year}
  {1975})}\BibitemShut {NoStop}%
\bibitem [{\citenamefont {Seul}\ and\ \citenamefont {Andelman}(1995)}]{Seul}%
  \BibitemOpen
  \bibfield  {author} {\bibinfo {author} {\bibfnamefont {M.}~\bibnamefont
  {Seul}}\ and\ \bibinfo {author} {\bibfnamefont {D.}~\bibnamefont
  {Andelman}},\ }\bibfield  {title} {\bibinfo {title} {Domain shapes and
  patterns: The phenomenology of modulated phases},\ }\href
  {https://doi.org/10.1126/science.267.5197.476} {\bibfield  {journal}
  {\bibinfo  {journal} {Science}\ }\textbf {\bibinfo {volume} {267}},\ \bibinfo
  {pages} {476} (\bibinfo {year} {1995})}\BibitemShut {NoStop}%
\bibitem [{\citenamefont {Chakrabarty}\ and\ \citenamefont
  {Nussinov}(2011)}]{modulation1}%
  \BibitemOpen
  \bibfield  {author} {\bibinfo {author} {\bibfnamefont {S.}~\bibnamefont
  {Chakrabarty}}\ and\ \bibinfo {author} {\bibfnamefont {Z.}~\bibnamefont
  {Nussinov}},\ }\bibfield  {title} {\bibinfo {title} {Modulation and
  correlations lengths in systems with competing interactions: new exponents
  and other universal features},\ }\href@noop {} {\bibfield  {journal}
  {\bibinfo  {journal} {Physical Review B}\ }\textbf {\bibinfo {volume} {84}},\
  \bibinfo {pages} {144402} (\bibinfo {year} {2011})}\BibitemShut {NoStop}%
\bibitem [{\citenamefont {Chakrabarty}\ \emph {et~al.}(2012)\citenamefont
  {Chakrabarty}, \citenamefont {Dobrosavljevic}, \citenamefont {Seidel},\ and\
  \citenamefont {Nussinov}}]{modulation2}%
  \BibitemOpen
  \bibfield  {author} {\bibinfo {author} {\bibfnamefont {S.}~\bibnamefont
  {Chakrabarty}}, \bibinfo {author} {\bibfnamefont {V.}~\bibnamefont
  {Dobrosavljevic}}, \bibinfo {author} {\bibfnamefont {A.}~\bibnamefont
  {Seidel}},\ and\ \bibinfo {author} {\bibfnamefont {Z.}~\bibnamefont
  {Nussinov}},\ }\bibfield  {title} {\bibinfo {title} {Universality of
  modulation length (and time) exponents},\ }\href@noop {} {\bibfield
  {journal} {\bibinfo  {journal} {Physical Review E}\ }\textbf {\bibinfo
  {volume} {86}},\ \bibinfo {pages} {041132} (\bibinfo {year}
  {2012})}\BibitemShut {NoStop}%
\bibitem [{Note2()}]{Note2}%
  \BibitemOpen
  \bibinfo {note} {This also includes models of magnetic materials \cite
  {Seul,Yoshimori,Hornnerich,Selke}, elastic stressed systems \cite {Amnon},
  liquid crystals \cite {Elihu,LC2}, chemical mixtures and membranes \cite
  {Seul}, and general Ginzburg-Landau theories describing competing orders
  \cite {compete_GL}. In these and numerous other arenas, the interplay between
  the varying order gradient terms leads to low energy states which may obtain
  their minima at finite wave-vectors. The resulting moat-type structures
  underlie basic aspects of crystallization \cite {Braz,Braz2}, superconductors
  in external magnetic fields \cite {FFLO1,FFLO2,FFLO3}, quantum Hall systems
  \cite {qhstr2,QHE,fogler1,chalker,qhstr1, qhstr3,Pu2024}, and countless
  others.}\BibitemShut {Stop}%
\bibitem [{\citenamefont {Pu}\ \emph {et~al.}(2024)\citenamefont {Pu},
  \citenamefont {Balrma}, \citenamefont {Taylor}, \citenamefont {Fradkin},\
  and\ \citenamefont {Papic}}]{Pu2024}%
  \BibitemOpen
  \bibfield  {author} {\bibinfo {author} {\bibfnamefont {S.}~\bibnamefont
  {Pu}}, \bibinfo {author} {\bibfnamefont {A.~C.}\ \bibnamefont {Balrma}},
  \bibinfo {author} {\bibfnamefont {J.}~\bibnamefont {Taylor}}, \bibinfo
  {author} {\bibfnamefont {E.}~\bibnamefont {Fradkin}},\ and\ \bibinfo {author}
  {\bibfnamefont {Z.}~\bibnamefont {Papic}},\ }\bibfield  {title} {\bibinfo
  {title} {Microscopic model for fractional quantum hall nematics},\
  }\href@noop {} {\bibfield  {journal} {\bibinfo  {journal} {Phys. Rev. Lett.}\
  }\textbf {\bibinfo {volume} {132}},\ \bibinfo {pages} {236503} (\bibinfo
  {year} {2024})}\BibitemShut {NoStop}%
\bibitem [{\citenamefont {Sedrakyan}\ \emph {et~al.}(2014)\citenamefont
  {Sedrakyan}, \citenamefont {Glazman},\ and\ \citenamefont
  {Kamenev}}]{Sedrakyan14}%
  \BibitemOpen
  \bibfield  {author} {\bibinfo {author} {\bibfnamefont {T.~A.}\ \bibnamefont
  {Sedrakyan}}, \bibinfo {author} {\bibfnamefont {L.~I.}\ \bibnamefont
  {Glazman}},\ and\ \bibinfo {author} {\bibfnamefont {A.}~\bibnamefont
  {Kamenev}},\ }\bibfield  {title} {\bibinfo {title} {Absence of bose
  condensation on lattices with moat bands},\ }\href@noop {} {\bibfield
  {journal} {\bibinfo  {journal} {Phys. Rev. B}\ }\textbf {\bibinfo {volume}
  {89}},\ \bibinfo {pages} {201112(R)} (\bibinfo {year} {2014})}\BibitemShut
  {NoStop}%
\bibitem [{\citenamefont {Meisinger}\ and\ \citenamefont
  {Ogilvie}(2013)}]{Meisinger:2012va}%
  \BibitemOpen
  \bibfield  {author} {\bibinfo {author} {\bibfnamefont {P.~N.}\ \bibnamefont
  {Meisinger}}\ and\ \bibinfo {author} {\bibfnamefont {M.~C.}\ \bibnamefont
  {Ogilvie}},\ }\bibfield  {title} {\bibinfo {title} {{PT Symmetry in Classical
  and Quantum Statistical Mechanics}},\ }\href
  {https://doi.org/10.1098/rsta.2012.0058} {\bibfield  {journal} {\bibinfo
  {journal} {Phil. Trans. Roy. Soc. Lond. A}\ }\textbf {\bibinfo {volume}
  {371}},\ \bibinfo {pages} {20120058} (\bibinfo {year} {2013})},\ \Eprint
  {https://arxiv.org/abs/1208.5077} {arXiv:1208.5077 [math-ph]} \BibitemShut
  {NoStop}%
\bibitem [{\citenamefont {Bender}\ and\ \citenamefont
  {Boettcher}(1998)}]{Bender:1998ke}%
  \BibitemOpen
  \bibfield  {author} {\bibinfo {author} {\bibfnamefont {C.~M.}\ \bibnamefont
  {Bender}}\ and\ \bibinfo {author} {\bibfnamefont {S.}~\bibnamefont
  {Boettcher}},\ }\bibfield  {title} {\bibinfo {title} {{Real spectra in
  nonHermitian Hamiltonians having PT symmetry}},\ }\href
  {https://doi.org/10.1103/PhysRevLett.80.5243} {\bibfield  {journal} {\bibinfo
   {journal} {Phys. Rev. Lett.}\ }\textbf {\bibinfo {volume} {80}},\ \bibinfo
  {pages} {5243} (\bibinfo {year} {1998})},\ \Eprint
  {https://arxiv.org/abs/physics/9712001} {arXiv:physics/9712001} \BibitemShut
  {NoStop}%
\bibitem [{\citenamefont {Bender}\ \emph {et~al.}(2019)\citenamefont {Bender},
  \citenamefont {Dorey}, \citenamefont {Dunning}, \citenamefont {Fring},
  \citenamefont {Hook}, \citenamefont {Jones}, \citenamefont {Kuzhel},
  \citenamefont {L\'evai},\ and\ \citenamefont {Tateo}}]{Bender:2019cwm}%
  \BibitemOpen
  \bibfield  {author} {\bibinfo {author} {\bibfnamefont {C.~M.}\ \bibnamefont
  {Bender}}, \bibinfo {author} {\bibfnamefont {P.~E.}\ \bibnamefont {Dorey}},
  \bibinfo {author} {\bibfnamefont {C.}~\bibnamefont {Dunning}}, \bibinfo
  {author} {\bibfnamefont {A.}~\bibnamefont {Fring}}, \bibinfo {author}
  {\bibfnamefont {D.~W.}\ \bibnamefont {Hook}}, \bibinfo {author}
  {\bibfnamefont {H.~F.}\ \bibnamefont {Jones}}, \bibinfo {author}
  {\bibfnamefont {S.}~\bibnamefont {Kuzhel}}, \bibinfo {author} {\bibfnamefont
  {G.}~\bibnamefont {L\'evai}},\ and\ \bibinfo {author} {\bibfnamefont
  {R.}~\bibnamefont {Tateo}},\ }\href {https://doi.org/10.1142/q0178} {\emph
  {\bibinfo {title} {{PT Symmetry}}}}\ (\bibinfo  {publisher} {WSP},\ \bibinfo
  {year} {2019})\BibitemShut {NoStop}%
\bibitem [{\citenamefont {Bender}\ and\ \citenamefont
  {Hook}(2023)}]{Bender:2023cem}%
  \BibitemOpen
  \bibfield  {author} {\bibinfo {author} {\bibfnamefont {C.~M.}\ \bibnamefont
  {Bender}}\ and\ \bibinfo {author} {\bibfnamefont {D.~W.}\ \bibnamefont
  {Hook}},\ }\bibfield  {title} {\bibinfo {title} {{PT-symmetric quantum
  mechanics}},\ }\href@noop {} {\  (\bibinfo {year} {2023})},\ \Eprint
  {https://arxiv.org/abs/2312.17386} {arXiv:2312.17386 [quant-ph]} \BibitemShut
  {NoStop}%
\bibitem [{\citenamefont {Thies}\ and\ \citenamefont
  {Urlichs}(2003)}]{Thies:2003kk}%
  \BibitemOpen
  \bibfield  {author} {\bibinfo {author} {\bibfnamefont {M.}~\bibnamefont
  {Thies}}\ and\ \bibinfo {author} {\bibfnamefont {K.}~\bibnamefont
  {Urlichs}},\ }\bibfield  {title} {\bibinfo {title} {{Revised phase diagram of
  the Gross-Neveu model}},\ }\href {https://doi.org/10.1103/PhysRevD.67.125015}
  {\bibfield  {journal} {\bibinfo  {journal} {Phys. Rev. D}\ }\textbf {\bibinfo
  {volume} {67}},\ \bibinfo {pages} {125015} (\bibinfo {year} {2003})},\
  \Eprint {https://arxiv.org/abs/hep-th/0302092} {arXiv:hep-th/0302092}
  \BibitemShut {NoStop}%
\bibitem [{\citenamefont {Basar}\ and\ \citenamefont
  {Dunne}(2008)}]{Basar:2008ki}%
  \BibitemOpen
  \bibfield  {author} {\bibinfo {author} {\bibfnamefont {G.}~\bibnamefont
  {Basar}}\ and\ \bibinfo {author} {\bibfnamefont {G.~V.}\ \bibnamefont
  {Dunne}},\ }\bibfield  {title} {\bibinfo {title} {{A Twisted Kink Crystal in
  the Chiral Gross-Neveu model}},\ }\href
  {https://doi.org/10.1103/PhysRevD.78.065022} {\bibfield  {journal} {\bibinfo
  {journal} {Phys. Rev. D}\ }\textbf {\bibinfo {volume} {78}},\ \bibinfo
  {pages} {065022} (\bibinfo {year} {2008})},\ \Eprint
  {https://arxiv.org/abs/0806.2659} {arXiv:0806.2659 [hep-th]} \BibitemShut
  {NoStop}%
\bibitem [{\citenamefont {Basar}\ \emph {et~al.}(2009)\citenamefont {Basar},
  \citenamefont {Dunne},\ and\ \citenamefont {Thies}}]{Basar:2009fg}%
  \BibitemOpen
  \bibfield  {author} {\bibinfo {author} {\bibfnamefont {G.}~\bibnamefont
  {Basar}}, \bibinfo {author} {\bibfnamefont {G.~V.}\ \bibnamefont {Dunne}},\
  and\ \bibinfo {author} {\bibfnamefont {M.}~\bibnamefont {Thies}},\ }\bibfield
   {title} {\bibinfo {title} {{Inhomogeneous Condensates in the Thermodynamics
  of the Chiral NJL(2) model}},\ }\href
  {https://doi.org/10.1103/PhysRevD.79.105012} {\bibfield  {journal} {\bibinfo
  {journal} {Phys. Rev.}\ }\textbf {\bibinfo {volume} {D79}},\ \bibinfo {pages}
  {105012} (\bibinfo {year} {2009})},\ \Eprint
  {https://arxiv.org/abs/0903.1868} {arXiv:0903.1868 [hep-th]} \BibitemShut
  {NoStop}%
%%CITATION = ARXIV:0903.1868;%%
\bibitem [{\citenamefont {Lenz}\ \emph {et~al.}(2020)\citenamefont {Lenz},
  \citenamefont {Pannullo}, \citenamefont {Wagner}, \citenamefont
  {Wellegehausen},\ and\ \citenamefont {Wipf}}]{Lenz:2020bxk}%
  \BibitemOpen
  \bibfield  {author} {\bibinfo {author} {\bibfnamefont {J.}~\bibnamefont
  {Lenz}}, \bibinfo {author} {\bibfnamefont {L.}~\bibnamefont {Pannullo}},
  \bibinfo {author} {\bibfnamefont {M.}~\bibnamefont {Wagner}}, \bibinfo
  {author} {\bibfnamefont {B.}~\bibnamefont {Wellegehausen}},\ and\ \bibinfo
  {author} {\bibfnamefont {A.}~\bibnamefont {Wipf}},\ }\bibfield  {title}
  {\bibinfo {title} {{Inhomogeneous phases in the Gross-Neveu model in 1+1
  dimensions at finite number of flavors}},\ }\href
  {https://doi.org/10.1103/PhysRevD.101.094512} {\bibfield  {journal} {\bibinfo
   {journal} {Phys. Rev. D}\ }\textbf {\bibinfo {volume} {101}},\ \bibinfo
  {pages} {094512} (\bibinfo {year} {2020})},\ \Eprint
  {https://arxiv.org/abs/2004.00295} {arXiv:2004.00295 [hep-lat]} \BibitemShut
  {NoStop}%
\bibitem [{\citenamefont {Lenz}\ \emph {et~al.}(2022)\citenamefont {Lenz},
  \citenamefont {Mandl},\ and\ \citenamefont {Wipf}}]{Lenz:2021kzo}%
  \BibitemOpen
  \bibfield  {author} {\bibinfo {author} {\bibfnamefont {J.~J.}\ \bibnamefont
  {Lenz}}, \bibinfo {author} {\bibfnamefont {M.}~\bibnamefont {Mandl}},\ and\
  \bibinfo {author} {\bibfnamefont {A.}~\bibnamefont {Wipf}},\ }\bibfield
  {title} {\bibinfo {title} {{Inhomogeneities in the two-flavor chiral
  Gross-Neveu model}},\ }\href {https://doi.org/10.1103/PhysRevD.105.034512}
  {\bibfield  {journal} {\bibinfo  {journal} {Phys. Rev. D}\ }\textbf {\bibinfo
  {volume} {105}},\ \bibinfo {pages} {034512} (\bibinfo {year} {2022})},\
  \Eprint {https://arxiv.org/abs/2109.05525} {arXiv:2109.05525 [hep-lat]}
  \BibitemShut {NoStop}%
\bibitem [{\citenamefont {Lenz}\ and\ \citenamefont
  {Mandl}(2022)}]{Lenz:2021vdz}%
  \BibitemOpen
  \bibfield  {author} {\bibinfo {author} {\bibfnamefont {J.~J.}\ \bibnamefont
  {Lenz}}\ and\ \bibinfo {author} {\bibfnamefont {M.}~\bibnamefont {Mandl}},\
  }\bibfield  {title} {\bibinfo {title} {{Remnants of large-$N_\mathrm{f}$
  inhomogeneities in the 2-flavor chiral Gross-Neveu model}},\ }\href
  {https://doi.org/10.22323/1.396.0415} {\bibfield  {journal} {\bibinfo
  {journal} {PoS}\ }\textbf {\bibinfo {volume} {LATTICE2021}},\ \bibinfo
  {pages} {415} (\bibinfo {year} {2022})},\ \Eprint
  {https://arxiv.org/abs/2110.12757} {arXiv:2110.12757 [hep-lat]} \BibitemShut
  {NoStop}%
\bibitem [{\citenamefont {Nonaka}\ and\ \citenamefont
  {Horie}(2022)}]{Nonaka:2021pwm}%
  \BibitemOpen
  \bibfield  {author} {\bibinfo {author} {\bibfnamefont {C.}~\bibnamefont
  {Nonaka}}\ and\ \bibinfo {author} {\bibfnamefont {K.}~\bibnamefont {Horie}},\
  }\bibfield  {title} {\bibinfo {title} {{Inhomogeneous phases in the chiral
  Gross-Neveu model on the lattice}},\ }\href
  {https://doi.org/10.22323/1.396.0150} {\bibfield  {journal} {\bibinfo
  {journal} {PoS}\ }\textbf {\bibinfo {volume} {LATTICE2021}},\ \bibinfo
  {pages} {150} (\bibinfo {year} {2022})},\ \Eprint
  {https://arxiv.org/abs/2112.02261} {arXiv:2112.02261 [hep-lat]} \BibitemShut
  {NoStop}%
\bibitem [{\citenamefont {Pannullo}(2023)}]{Pannullo:2023cat}%
  \BibitemOpen
  \bibfield  {author} {\bibinfo {author} {\bibfnamefont {L.}~\bibnamefont
  {Pannullo}},\ }\bibfield  {title} {\bibinfo {title} {{Inhomogeneous
  condensation in the Gross-Neveu model in noninteger spatial dimensions
  1\ensuremath{\leq}d\ensuremath{<}3}},\ }\href
  {https://doi.org/10.1103/PhysRevD.108.036022} {\bibfield  {journal} {\bibinfo
   {journal} {Phys. Rev. D}\ }\textbf {\bibinfo {volume} {108}},\ \bibinfo
  {pages} {036022} (\bibinfo {year} {2023})},\ \Eprint
  {https://arxiv.org/abs/2306.16290} {arXiv:2306.16290 [hep-ph]} \BibitemShut
  {NoStop}%
\bibitem [{\citenamefont {Koenigstein}\ and\ \citenamefont
  {Pannullo}(2024)}]{Koenigstein:2023yzv}%
  \BibitemOpen
  \bibfield  {author} {\bibinfo {author} {\bibfnamefont {A.}~\bibnamefont
  {Koenigstein}}\ and\ \bibinfo {author} {\bibfnamefont {L.}~\bibnamefont
  {Pannullo}},\ }\bibfield  {title} {\bibinfo {title} {{Inhomogeneous
  condensation in the Gross-Neveu model in noninteger spatial dimensions
  1\ensuremath{\leq}d\ensuremath{<}3. II. Nonzero temperature and chemical
  potential}},\ }\href {https://doi.org/10.1103/PhysRevD.109.056015} {\bibfield
   {journal} {\bibinfo  {journal} {Phys. Rev. D}\ }\textbf {\bibinfo {volume}
  {109}},\ \bibinfo {pages} {056015} (\bibinfo {year} {2024})},\ \Eprint
  {https://arxiv.org/abs/2312.04904} {arXiv:2312.04904 [hep-ph]} \BibitemShut
  {NoStop}%
\bibitem [{\citenamefont {Koenigstein}\ \emph {et~al.}(2022)\citenamefont
  {Koenigstein}, \citenamefont {Pannullo}, \citenamefont {Rechenberger},
  \citenamefont {Steil},\ and\ \citenamefont {Winstel}}]{Koenigstein:2021llr}%
  \BibitemOpen
  \bibfield  {author} {\bibinfo {author} {\bibfnamefont {A.}~\bibnamefont
  {Koenigstein}}, \bibinfo {author} {\bibfnamefont {L.}~\bibnamefont
  {Pannullo}}, \bibinfo {author} {\bibfnamefont {S.}~\bibnamefont
  {Rechenberger}}, \bibinfo {author} {\bibfnamefont {M.~J.}\ \bibnamefont
  {Steil}},\ and\ \bibinfo {author} {\bibfnamefont {M.}~\bibnamefont
  {Winstel}},\ }\bibfield  {title} {\bibinfo {title} {{Detecting inhomogeneous
  chiral condensation from the bosonic two-point function in the (1 +
  1)-dimensional Gross\textendash{}Neveu model in the mean-field
  approximation*}},\ }\href {https://doi.org/10.1088/1751-8121/ac820a}
  {\bibfield  {journal} {\bibinfo  {journal} {J. Phys. A}\ }\textbf {\bibinfo
  {volume} {55}},\ \bibinfo {pages} {375402} (\bibinfo {year} {2022})},\
  \Eprint {https://arxiv.org/abs/2112.07024} {arXiv:2112.07024 [hep-ph]}
  \BibitemShut {NoStop}%
\bibitem [{\citenamefont {Nickel}(2009)}]{Nickel:2009wj}%
  \BibitemOpen
  \bibfield  {author} {\bibinfo {author} {\bibfnamefont {D.}~\bibnamefont
  {Nickel}},\ }\bibfield  {title} {\bibinfo {title} {Inhomogeneous phases in
  the {{Nambu-Jona-Lasino}} and quark-meson model},\ }\href
  {https://doi.org/10.1103/PhysRevD.80.074025} {\bibfield  {journal} {\bibinfo
  {journal} {Physical Review D}\ }\textbf {\bibinfo {volume} {80}},\ \bibinfo
  {pages} {074025} (\bibinfo {year} {2009})},\ \Eprint
  {https://arxiv.org/abs/0906.5295} {arxiv:0906.5295} \BibitemShut {NoStop}%
\bibitem [{\citenamefont {Buballa}\ and\ \citenamefont
  {Carignano}(2016)}]{Buballa:2015awa}%
  \BibitemOpen
  \bibfield  {author} {\bibinfo {author} {\bibfnamefont {M.}~\bibnamefont
  {Buballa}}\ and\ \bibinfo {author} {\bibfnamefont {S.}~\bibnamefont
  {Carignano}},\ }\bibfield  {title} {\bibinfo {title} {{Inhomogeneous chiral
  symmetry breaking in dense neutron-star matter}},\ }\href
  {https://doi.org/10.1140/epja/i2016-16057-6} {\bibfield  {journal} {\bibinfo
  {journal} {Eur. Phys. J.}\ }\textbf {\bibinfo {volume} {A52}},\ \bibinfo
  {pages} {57} (\bibinfo {year} {2016})},\ \Eprint
  {https://arxiv.org/abs/1508.04361} {arXiv:1508.04361 [nucl-th]} \BibitemShut
  {NoStop}%
%%CITATION = ARXIV:1508.04361;%%
\bibitem [{\citenamefont {Lee}\ \emph {et~al.}(2015)\citenamefont {Lee},
  \citenamefont {Nakano}, \citenamefont {Tsue}, \citenamefont {Tatsumi},\ and\
  \citenamefont {Friman}}]{Lee:2015bva}%
  \BibitemOpen
  \bibfield  {author} {\bibinfo {author} {\bibfnamefont {T.-G.}\ \bibnamefont
  {Lee}}, \bibinfo {author} {\bibfnamefont {E.}~\bibnamefont {Nakano}},
  \bibinfo {author} {\bibfnamefont {Y.}~\bibnamefont {Tsue}}, \bibinfo {author}
  {\bibfnamefont {T.}~\bibnamefont {Tatsumi}},\ and\ \bibinfo {author}
  {\bibfnamefont {B.}~\bibnamefont {Friman}},\ }\bibfield  {title} {\bibinfo
  {title} {{Landau-Peierls instability in a Fulde-Ferrell type inhomogeneous
  chiral condensed phase}},\ }\href
  {https://doi.org/10.1103/PhysRevD.92.034024} {\bibfield  {journal} {\bibinfo
  {journal} {Phys. Rev.}\ }\textbf {\bibinfo {volume} {D92}},\ \bibinfo {pages}
  {034024} (\bibinfo {year} {2015})},\ \Eprint
  {https://arxiv.org/abs/1504.03185} {arXiv:1504.03185 [hep-ph]} \BibitemShut
  {NoStop}%
%%CITATION = ARXIV:1504.03185;%%
\bibitem [{\citenamefont {Hidaka}\ \emph {et~al.}(2015)\citenamefont {Hidaka},
  \citenamefont {Kamikado}, \citenamefont {Kanazawa},\ and\ \citenamefont
  {Noumi}}]{Hidaka:2015xza}%
  \BibitemOpen
  \bibfield  {author} {\bibinfo {author} {\bibfnamefont {Y.}~\bibnamefont
  {Hidaka}}, \bibinfo {author} {\bibfnamefont {K.}~\bibnamefont {Kamikado}},
  \bibinfo {author} {\bibfnamefont {T.}~\bibnamefont {Kanazawa}},\ and\
  \bibinfo {author} {\bibfnamefont {T.}~\bibnamefont {Noumi}},\ }\bibfield
  {title} {\bibinfo {title} {{Phonons, pions and quasi-long-range order in
  spatially modulated chiral condensates}},\ }\href
  {https://doi.org/10.1103/PhysRevD.92.034003} {\bibfield  {journal} {\bibinfo
  {journal} {Phys. Rev.}\ }\textbf {\bibinfo {volume} {D92}},\ \bibinfo {pages}
  {034003} (\bibinfo {year} {2015})},\ \Eprint
  {https://arxiv.org/abs/1505.00848} {arXiv:1505.00848 [hep-ph]} \BibitemShut
  {NoStop}%
%%CITATION = ARXIV:1505.00848;%%
\bibitem [{\citenamefont {Buballa}\ and\ \citenamefont
  {Carignano}(2019)}]{Buballa:2018hux}%
  \BibitemOpen
  \bibfield  {author} {\bibinfo {author} {\bibfnamefont {M.}~\bibnamefont
  {Buballa}}\ and\ \bibinfo {author} {\bibfnamefont {S.}~\bibnamefont
  {Carignano}},\ }\bibfield  {title} {\bibinfo {title} {{Inhomogeneous chiral
  phases away from the chiral limit}},\ }\href
  {https://doi.org/10.1016/j.physletb.2019.02.045} {\bibfield  {journal}
  {\bibinfo  {journal} {Phys. Lett. B}\ }\textbf {\bibinfo {volume} {791}},\
  \bibinfo {pages} {361} (\bibinfo {year} {2019})},\ \Eprint
  {https://arxiv.org/abs/1809.10066} {arXiv:1809.10066 [hep-ph]} \BibitemShut
  {NoStop}%
\bibitem [{\citenamefont {Carignano}\ and\ \citenamefont
  {Buballa}(2020)}]{Carignano:2019ivp}%
  \BibitemOpen
  \bibfield  {author} {\bibinfo {author} {\bibfnamefont {S.}~\bibnamefont
  {Carignano}}\ and\ \bibinfo {author} {\bibfnamefont {M.}~\bibnamefont
  {Buballa}},\ }\bibfield  {title} {\bibinfo {title} {{Inhomogeneous chiral
  condensates in three-flavor quark matter}},\ }\href
  {https://doi.org/10.1103/PhysRevD.101.014026} {\bibfield  {journal} {\bibinfo
   {journal} {Phys. Rev. D}\ }\textbf {\bibinfo {volume} {101}},\ \bibinfo
  {pages} {014026} (\bibinfo {year} {2020})},\ \Eprint
  {https://arxiv.org/abs/1910.03604} {arXiv:1910.03604 [hep-ph]} \BibitemShut
  {NoStop}%
\bibitem [{\citenamefont {Pannullo}\ \emph {et~al.}(2024)\citenamefont
  {Pannullo}, \citenamefont {Wagner},\ and\ \citenamefont
  {Winstel}}]{Pannullo:2024sov}%
  \BibitemOpen
  \bibfield  {author} {\bibinfo {author} {\bibfnamefont {L.}~\bibnamefont
  {Pannullo}}, \bibinfo {author} {\bibfnamefont {M.}~\bibnamefont {Wagner}},\
  and\ \bibinfo {author} {\bibfnamefont {M.}~\bibnamefont {Winstel}},\
  }\bibfield  {title} {\bibinfo {title} {{Regularization effects in the
  Nambu-Jona-Lasinio model: Strong scheme dependence of inhomogeneous phases
  and persistence of the moat regime}},\ }\href@noop {} {\  (\bibinfo {year}
  {2024})},\ \Eprint {https://arxiv.org/abs/2406.11312} {arXiv:2406.11312
  [hep-ph]} \BibitemShut {NoStop}%
\bibitem [{\citenamefont {Motta}\ \emph
  {et~al.}(2024{\natexlab{a}})\citenamefont {Motta}, \citenamefont {Bernhardt},
  \citenamefont {Buballa},\ and\ \citenamefont {Fischer}}]{Motta:2024agi}%
  \BibitemOpen
  \bibfield  {author} {\bibinfo {author} {\bibfnamefont {T.~F.}\ \bibnamefont
  {Motta}}, \bibinfo {author} {\bibfnamefont {J.}~\bibnamefont {Bernhardt}},
  \bibinfo {author} {\bibfnamefont {M.}~\bibnamefont {Buballa}},\ and\ \bibinfo
  {author} {\bibfnamefont {C.~S.}\ \bibnamefont {Fischer}},\ }\bibfield
  {title} {\bibinfo {title} {{Inhomogenuous instabilities at large chemical
  potential in a rainbow-ladder QCD model}},\ }\href@noop {} {\  (\bibinfo
  {year} {2024}{\natexlab{a}})},\ \Eprint {https://arxiv.org/abs/2406.00205}
  {arXiv:2406.00205 [hep-ph]} \BibitemShut {NoStop}%
\bibitem [{\citenamefont {Motta}\ \emph
  {et~al.}(2024{\natexlab{b}})\citenamefont {Motta}, \citenamefont {Buballa},\
  and\ \citenamefont {Fischer}}]{Motta:2024rvk}%
  \BibitemOpen
  \bibfield  {author} {\bibinfo {author} {\bibfnamefont {T.~F.}\ \bibnamefont
  {Motta}}, \bibinfo {author} {\bibfnamefont {M.}~\bibnamefont {Buballa}},\
  and\ \bibinfo {author} {\bibfnamefont {C.~S.}\ \bibnamefont {Fischer}},\
  }\bibfield  {title} {\bibinfo {title} {{New Tool to Detect Inhomogeneous
  Chiral Symmetry Breaking}},\ }\href@noop {} {\  (\bibinfo {year}
  {2024}{\natexlab{b}})},\ \Eprint {https://arxiv.org/abs/2411.02285}
  {arXiv:2411.02285 [hep-ph]} \BibitemShut {NoStop}%
\bibitem [{\citenamefont {Nishimura}\ \emph {et~al.}(2014)\citenamefont
  {Nishimura}, \citenamefont {Ogilvie},\ and\ \citenamefont
  {Pangeni}}]{Nishimura:2014rxa}%
  \BibitemOpen
  \bibfield  {author} {\bibinfo {author} {\bibfnamefont {H.}~\bibnamefont
  {Nishimura}}, \bibinfo {author} {\bibfnamefont {M.~C.}\ \bibnamefont
  {Ogilvie}},\ and\ \bibinfo {author} {\bibfnamefont {K.}~\bibnamefont
  {Pangeni}},\ }\bibfield  {title} {\bibinfo {title} {{Complex saddle points in
  QCD at finite temperature and density}},\ }\href
  {https://doi.org/10.1103/PhysRevD.90.045039} {\bibfield  {journal} {\bibinfo
  {journal} {Phys. Rev. D}\ }\textbf {\bibinfo {volume} {90}},\ \bibinfo
  {pages} {045039} (\bibinfo {year} {2014})},\ \Eprint
  {https://arxiv.org/abs/1401.7982} {arXiv:1401.7982 [hep-ph]} \BibitemShut
  {NoStop}%
\bibitem [{\citenamefont {Nishimura}\ \emph {et~al.}(2015)\citenamefont
  {Nishimura}, \citenamefont {Ogilvie},\ and\ \citenamefont
  {Pangeni}}]{Nishimura:2014kla}%
  \BibitemOpen
  \bibfield  {author} {\bibinfo {author} {\bibfnamefont {H.}~\bibnamefont
  {Nishimura}}, \bibinfo {author} {\bibfnamefont {M.~C.}\ \bibnamefont
  {Ogilvie}},\ and\ \bibinfo {author} {\bibfnamefont {K.}~\bibnamefont
  {Pangeni}},\ }\bibfield  {title} {\bibinfo {title} {{Complex Saddle Points
  and Disorder Lines in QCD at finite temperature and density}},\ }\href
  {https://doi.org/10.1103/PhysRevD.91.054004} {\bibfield  {journal} {\bibinfo
  {journal} {Phys. Rev. D}\ }\textbf {\bibinfo {volume} {91}},\ \bibinfo
  {pages} {054004} (\bibinfo {year} {2015})},\ \Eprint
  {https://arxiv.org/abs/1411.4959} {arXiv:1411.4959 [hep-ph]} \BibitemShut
  {NoStop}%
\bibitem [{\citenamefont {Haensch}\ \emph {et~al.}(2024)\citenamefont
  {Haensch}, \citenamefont {Rennecke},\ and\ \citenamefont {von
  Smekal}}]{Haensch:2023sig}%
  \BibitemOpen
  \bibfield  {author} {\bibinfo {author} {\bibfnamefont {M.}~\bibnamefont
  {Haensch}}, \bibinfo {author} {\bibfnamefont {F.}~\bibnamefont {Rennecke}},\
  and\ \bibinfo {author} {\bibfnamefont {L.}~\bibnamefont {von Smekal}},\
  }\bibfield  {title} {\bibinfo {title} {{Medium induced mixing, spatial
  modulations, and critical modes in QCD}},\ }\href
  {https://doi.org/10.1103/PhysRevD.110.036018} {\bibfield  {journal} {\bibinfo
   {journal} {Phys. Rev. D}\ }\textbf {\bibinfo {volume} {110}},\ \bibinfo
  {pages} {036018} (\bibinfo {year} {2024})},\ \Eprint
  {https://arxiv.org/abs/2308.16244} {arXiv:2308.16244 [hep-ph]} \BibitemShut
  {NoStop}%
\bibitem [{\citenamefont {T\"opfel}\ \emph {et~al.}(2024)\citenamefont
  {T\"opfel}, \citenamefont {Pawlowski},\ and\ \citenamefont
  {Braun}}]{Topfel:2024iop}%
  \BibitemOpen
  \bibfield  {author} {\bibinfo {author} {\bibfnamefont {S.}~\bibnamefont
  {T\"opfel}}, \bibinfo {author} {\bibfnamefont {J.~M.}\ \bibnamefont
  {Pawlowski}},\ and\ \bibinfo {author} {\bibfnamefont {J.}~\bibnamefont
  {Braun}},\ }\bibfield  {title} {\bibinfo {title} {{Phase structure of quark
  matter and in-medium properties of mesons from Callan-Symanzik flows}},\
  }\href@noop {} {\  (\bibinfo {year} {2024})},\ \Eprint
  {https://arxiv.org/abs/2412.16059} {arXiv:2412.16059 [hep-ph]} \BibitemShut
  {NoStop}%
\bibitem [{\citenamefont {Nishimura}\ \emph {et~al.}(2016)\citenamefont
  {Nishimura}, \citenamefont {Ogilvie},\ and\ \citenamefont
  {Pangeni}}]{Nishimura:2015lit}%
  \BibitemOpen
  \bibfield  {author} {\bibinfo {author} {\bibfnamefont {H.}~\bibnamefont
  {Nishimura}}, \bibinfo {author} {\bibfnamefont {M.~C.}\ \bibnamefont
  {Ogilvie}},\ and\ \bibinfo {author} {\bibfnamefont {K.}~\bibnamefont
  {Pangeni}},\ }\bibfield  {title} {\bibinfo {title} {{Complex spectrum of
  finite-density lattice QCD with static quarks at strong coupling}},\ }\href
  {https://doi.org/10.1103/PhysRevD.93.094501} {\bibfield  {journal} {\bibinfo
  {journal} {Phys. Rev. D}\ }\textbf {\bibinfo {volume} {93}},\ \bibinfo
  {pages} {094501} (\bibinfo {year} {2016})},\ \Eprint
  {https://arxiv.org/abs/1512.09131} {arXiv:1512.09131 [hep-lat]} \BibitemShut
  {NoStop}%
\bibitem [{\citenamefont {Nishimura}\ \emph {et~al.}(2017)\citenamefont
  {Nishimura}, \citenamefont {Ogilvie},\ and\ \citenamefont
  {Pangeni}}]{Nishimura:2016yue}%
  \BibitemOpen
  \bibfield  {author} {\bibinfo {author} {\bibfnamefont {H.}~\bibnamefont
  {Nishimura}}, \bibinfo {author} {\bibfnamefont {M.~C.}\ \bibnamefont
  {Ogilvie}},\ and\ \bibinfo {author} {\bibfnamefont {K.}~\bibnamefont
  {Pangeni}},\ }\bibfield  {title} {\bibinfo {title} {{Liquid-Gas Phase
  Transitions and $\mathcal{CK}$ Symmetry in Quantum Field Theories}},\ }\href
  {https://doi.org/10.1103/PhysRevD.95.076003} {\bibfield  {journal} {\bibinfo
  {journal} {Phys. Rev. D}\ }\textbf {\bibinfo {volume} {95}},\ \bibinfo
  {pages} {076003} (\bibinfo {year} {2017})},\ \Eprint
  {https://arxiv.org/abs/1612.09575} {arXiv:1612.09575 [hep-th]} \BibitemShut
  {NoStop}%
\bibitem [{\citenamefont {Winstel}(2024)}]{Winstel:2024dqu}%
  \BibitemOpen
  \bibfield  {author} {\bibinfo {author} {\bibfnamefont {M.}~\bibnamefont
  {Winstel}},\ }\bibfield  {title} {\bibinfo {title} {{Spatially oscillating
  correlation functions in (2+1)-dimensional four-fermion models: The mixing of
  scalar and vector modes at finite density}},\ }\href
  {https://doi.org/10.1103/PhysRevD.110.034008} {\bibfield  {journal} {\bibinfo
   {journal} {Phys. Rev. D}\ }\textbf {\bibinfo {volume} {110}},\ \bibinfo
  {pages} {034008} (\bibinfo {year} {2024})},\ \Eprint
  {https://arxiv.org/abs/2403.07430} {arXiv:2403.07430 [hep-ph]} \BibitemShut
  {NoStop}%
\bibitem [{Note3()}]{Note3}%
  \BibitemOpen
  \bibinfo {note} {In perturbative QCD with massless quarks, Friedel
  oscillations arise as an oscillatory function times a power law
  fall-off.\cite {Kapusta:1988fi}}\BibitemShut {NoStop}%
\bibitem [{\citenamefont {Pisarski}\ \emph {et~al.}(2019)\citenamefont
  {Pisarski}, \citenamefont {Skokov},\ and\ \citenamefont
  {Tsvelik}}]{Pisarski:2018bct}%
  \BibitemOpen
  \bibfield  {author} {\bibinfo {author} {\bibfnamefont {R.~D.}\ \bibnamefont
  {Pisarski}}, \bibinfo {author} {\bibfnamefont {V.~V.}\ \bibnamefont
  {Skokov}},\ and\ \bibinfo {author} {\bibfnamefont {A.~M.}\ \bibnamefont
  {Tsvelik}},\ }\bibfield  {title} {\bibinfo {title} {{Fluctuations in cool
  quark matter and the phase diagram of Quantum Chromodynamics}},\ }\href
  {https://doi.org/10.1103/PhysRevD.99.074025} {\bibfield  {journal} {\bibinfo
  {journal} {Phys. Rev. D}\ }\textbf {\bibinfo {volume} {99}},\ \bibinfo
  {pages} {074025} (\bibinfo {year} {2019})},\ \Eprint
  {https://arxiv.org/abs/1801.08156} {arXiv:1801.08156 [hep-ph]} \BibitemShut
  {NoStop}%
\bibitem [{\citenamefont {Pisarski}\ \emph {et~al.}(2020)\citenamefont
  {Pisarski}, \citenamefont {Tsvelik},\ and\ \citenamefont
  {Valgushev}}]{Pisarski:2020dnx}%
  \BibitemOpen
  \bibfield  {author} {\bibinfo {author} {\bibfnamefont {R.~D.}\ \bibnamefont
  {Pisarski}}, \bibinfo {author} {\bibfnamefont {A.~M.}\ \bibnamefont
  {Tsvelik}},\ and\ \bibinfo {author} {\bibfnamefont {S.}~\bibnamefont
  {Valgushev}},\ }\bibfield  {title} {\bibinfo {title} {{How transverse thermal
  fluctuations disorder a condensate of chiral spirals into a quantum spin
  liquid}},\ }\href {https://doi.org/10.1103/PhysRevD.102.016015} {\bibfield
  {journal} {\bibinfo  {journal} {Phys. Rev. D}\ }\textbf {\bibinfo {volume}
  {102}},\ \bibinfo {pages} {016015} (\bibinfo {year} {2020})},\ \Eprint
  {https://arxiv.org/abs/2005.10259} {arXiv:2005.10259 [hep-ph]} \BibitemShut
  {NoStop}%
\bibitem [{\citenamefont {Winstel}\ and\ \citenamefont
  {Valgushev}(2024)}]{Winstel:2024qle}%
  \BibitemOpen
  \bibfield  {author} {\bibinfo {author} {\bibfnamefont {M.}~\bibnamefont
  {Winstel}}\ and\ \bibinfo {author} {\bibfnamefont {S.}~\bibnamefont
  {Valgushev}},\ }\bibfield  {title} {\bibinfo {title} {{Lattice study of
  disordering of inhomogeneous condensates and the Quantum Pion Liquid in
  effective $O(N)$ model}},\ }in\ \href@noop {} {\emph {\bibinfo {booktitle}
  {{Excited QCD 2024 Workshop}}}}\ (\bibinfo {year} {2024})\ \Eprint
  {https://arxiv.org/abs/2403.18640} {arXiv:2403.18640 [hep-lat]} \BibitemShut
  {NoStop}%
\bibitem [{\citenamefont {Pisarski}\ \emph {et~al.}(2021)\citenamefont
  {Pisarski}, \citenamefont {Rennecke}, \citenamefont {Tsvelik},\ and\
  \citenamefont {Valgushev}}]{Pisarski:2020gkx}%
  \BibitemOpen
  \bibfield  {author} {\bibinfo {author} {\bibfnamefont {R.~D.}\ \bibnamefont
  {Pisarski}}, \bibinfo {author} {\bibfnamefont {F.}~\bibnamefont {Rennecke}},
  \bibinfo {author} {\bibfnamefont {A.}~\bibnamefont {Tsvelik}},\ and\ \bibinfo
  {author} {\bibfnamefont {S.}~\bibnamefont {Valgushev}},\ }\bibfield  {title}
  {\bibinfo {title} {{The Lifshitz Regime and its Experimental Signals}},\
  }\href {https://doi.org/10.1016/j.nuclphysa.2020.121910} {\bibfield
  {journal} {\bibinfo  {journal} {Nucl. Phys. A}\ }\textbf {\bibinfo {volume}
  {1005}},\ \bibinfo {pages} {121910} (\bibinfo {year} {2021})},\ \Eprint
  {https://arxiv.org/abs/2005.00045} {arXiv:2005.00045 [nucl-th]} \BibitemShut
  {NoStop}%
\bibitem [{\citenamefont {Pisarski}\ and\ \citenamefont
  {Rennecke}(2021)}]{Pisarski:2021qof}%
  \BibitemOpen
  \bibfield  {author} {\bibinfo {author} {\bibfnamefont {R.~D.}\ \bibnamefont
  {Pisarski}}\ and\ \bibinfo {author} {\bibfnamefont {F.}~\bibnamefont
  {Rennecke}},\ }\bibfield  {title} {\bibinfo {title} {{Signatures of Moat
  Regimes in Heavy-Ion Collisions}},\ }\href
  {https://doi.org/10.1103/PhysRevLett.127.152302} {\bibfield  {journal}
  {\bibinfo  {journal} {Phys. Rev. Lett.}\ }\textbf {\bibinfo {volume} {127}},\
  \bibinfo {pages} {152302} (\bibinfo {year} {2021})},\ \Eprint
  {https://arxiv.org/abs/2103.06890} {arXiv:2103.06890 [hep-ph]} \BibitemShut
  {NoStop}%
\bibitem [{\citenamefont {Rennecke}\ \emph {et~al.}(2023)\citenamefont
  {Rennecke}, \citenamefont {Pisarski},\ and\ \citenamefont
  {Rischke}}]{Rennecke:2023xhc}%
  \BibitemOpen
  \bibfield  {author} {\bibinfo {author} {\bibfnamefont {F.}~\bibnamefont
  {Rennecke}}, \bibinfo {author} {\bibfnamefont {R.~D.}\ \bibnamefont
  {Pisarski}},\ and\ \bibinfo {author} {\bibfnamefont {D.~H.}\ \bibnamefont
  {Rischke}},\ }\bibfield  {title} {\bibinfo {title} {{Particle interferometry
  in a moat regime}},\ }\href {https://doi.org/10.1103/PhysRevD.107.116011}
  {\bibfield  {journal} {\bibinfo  {journal} {Phys. Rev. D}\ }\textbf {\bibinfo
  {volume} {107}},\ \bibinfo {pages} {116011} (\bibinfo {year} {2023})},\
  \Eprint {https://arxiv.org/abs/2301.11484} {arXiv:2301.11484 [hep-ph]}
  \BibitemShut {NoStop}%
\bibitem [{\citenamefont {Fukushima}\ \emph {et~al.}(2024)\citenamefont
  {Fukushima}, \citenamefont {Hidaka}, \citenamefont {Inoue}, \citenamefont
  {Shigaki},\ and\ \citenamefont {Yamaguchi}}]{Fukushima:2023tpv}%
  \BibitemOpen
  \bibfield  {author} {\bibinfo {author} {\bibfnamefont {K.}~\bibnamefont
  {Fukushima}}, \bibinfo {author} {\bibfnamefont {Y.}~\bibnamefont {Hidaka}},
  \bibinfo {author} {\bibfnamefont {K.}~\bibnamefont {Inoue}}, \bibinfo
  {author} {\bibfnamefont {K.}~\bibnamefont {Shigaki}},\ and\ \bibinfo {author}
  {\bibfnamefont {Y.}~\bibnamefont {Yamaguchi}},\ }\bibfield  {title} {\bibinfo
  {title} {{Hanbury-Brown\textendash{}Twiss signature for clustered
  substructures probing primordial inhomogeneity in hot and dense QCD
  matter}},\ }\href {https://doi.org/10.1103/PhysRevC.109.L051903} {\bibfield
  {journal} {\bibinfo  {journal} {Phys. Rev. C}\ }\textbf {\bibinfo {volume}
  {109}},\ \bibinfo {pages} {L051903} (\bibinfo {year} {2024})},\ \Eprint
  {https://arxiv.org/abs/2306.17619} {arXiv:2306.17619 [hep-ph]} \BibitemShut
  {NoStop}%
\bibitem [{\citenamefont {Nishimura}\ \emph {et~al.}(2022)\citenamefont
  {Nishimura}, \citenamefont {Kitazawa},\ and\ \citenamefont
  {Kunihiro}}]{Nishimura:2022mku}%
  \BibitemOpen
  \bibfield  {author} {\bibinfo {author} {\bibfnamefont {T.}~\bibnamefont
  {Nishimura}}, \bibinfo {author} {\bibfnamefont {M.}~\bibnamefont
  {Kitazawa}},\ and\ \bibinfo {author} {\bibfnamefont {T.}~\bibnamefont
  {Kunihiro}},\ }\bibfield  {title} {\bibinfo {title} {{Anomalous enhancement
  of dilepton production as a precursor of color superconductivity}},\ }\href
  {https://doi.org/10.1093/ptep/ptac100} {\bibfield  {journal} {\bibinfo
  {journal} {PTEP}\ }\textbf {\bibinfo {volume} {2022}},\ \bibinfo {pages}
  {093D02} (\bibinfo {year} {2022})},\ \Eprint
  {https://arxiv.org/abs/2201.01963} {arXiv:2201.01963 [hep-ph]} \BibitemShut
  {NoStop}%
\bibitem [{\citenamefont {Nishimura}\ \emph {et~al.}(2023)\citenamefont
  {Nishimura}, \citenamefont {Kitazawa},\ and\ \citenamefont
  {Kunihiro}}]{Nishimura:2023oqn}%
  \BibitemOpen
  \bibfield  {author} {\bibinfo {author} {\bibfnamefont {T.}~\bibnamefont
  {Nishimura}}, \bibinfo {author} {\bibfnamefont {M.}~\bibnamefont
  {Kitazawa}},\ and\ \bibinfo {author} {\bibfnamefont {T.}~\bibnamefont
  {Kunihiro}},\ }\bibfield  {title} {\bibinfo {title} {{Enhancement of dilepton
  production rate and electric conductivity around the QCD critical point}},\
  }\href {https://doi.org/10.1093/ptep/ptad051} {\bibfield  {journal} {\bibinfo
   {journal} {PTEP}\ }\textbf {\bibinfo {volume} {2023}},\ \bibinfo {pages}
  {053D01} (\bibinfo {year} {2023})},\ \Eprint
  {https://arxiv.org/abs/2302.03191} {arXiv:2302.03191 [hep-ph]} \BibitemShut
  {NoStop}%
\bibitem [{\citenamefont {Nishimura}\ \emph {et~al.}(2024)\citenamefont
  {Nishimura}, \citenamefont {Kitazawa},\ and\ \citenamefont
  {Kunihiro}}]{Nishimura:2024kvz}%
  \BibitemOpen
  \bibfield  {author} {\bibinfo {author} {\bibfnamefont {T.}~\bibnamefont
  {Nishimura}}, \bibinfo {author} {\bibfnamefont {M.}~\bibnamefont
  {Kitazawa}},\ and\ \bibinfo {author} {\bibfnamefont {T.}~\bibnamefont
  {Kunihiro}},\ }\bibfield  {title} {\bibinfo {title} {{Electromagnetic
  response of dense quark matter around color-superconducting phase transition
  and QCD critical point}},\ }\href {https://doi.org/10.1016/j.aop.2024.169768}
  {\bibfield  {journal} {\bibinfo  {journal} {Annals Phys.}\ }\textbf {\bibinfo
  {volume} {469}},\ \bibinfo {pages} {169768} (\bibinfo {year} {2024})},\
  \Eprint {https://arxiv.org/abs/2405.09240} {arXiv:2405.09240 [hep-ph]}
  \BibitemShut {NoStop}%
\bibitem [{\citenamefont {Hayashi}\ and\ \citenamefont
  {Tsue}(2024)}]{Hayashi:2024sae}%
  \BibitemOpen
  \bibfield  {author} {\bibinfo {author} {\bibfnamefont {K.}~\bibnamefont
  {Hayashi}}\ and\ \bibinfo {author} {\bibfnamefont {Y.}~\bibnamefont {Tsue}},\
  }\bibfield  {title} {\bibinfo {title} {{Modified dilepton production rate
  from charged pion-pair annihilation in the inhomogeneous chiral condensed
  phase}},\ }\href@noop {} {\  (\bibinfo {year} {2024})},\ \Eprint
  {https://arxiv.org/abs/2407.08523} {arXiv:2407.08523 [hep-ph]} \BibitemShut
  {NoStop}%
\bibitem [{Note4()}]{Note4}%
  \BibitemOpen
  \bibinfo {note} {As this work was in preparation, Ref.~\cite
  {Hayashi:2024sae} appeared, and analyzed dilepton production from a specific
  inhomogeneous phase. Ref. \cite {Hayashi:2024sae} includes the modified
  dispersion relation of a dual chiral density wave, but not the changes to the
  photon-moaton vertices. This is discussed in the Supplemental
  Material.}\BibitemShut {Stop}%
\bibitem [{\citenamefont {Balog}\ \emph {et~al.}(2019)\citenamefont {Balog},
  \citenamefont {Chat\'e}, \citenamefont {Delamotte}, \citenamefont
  {Marohnic},\ and\ \citenamefont {Wschebor}}]{Balog:2019rrg}%
  \BibitemOpen
  \bibfield  {author} {\bibinfo {author} {\bibfnamefont {I.}~\bibnamefont
  {Balog}}, \bibinfo {author} {\bibfnamefont {H.}~\bibnamefont {Chat\'e}},
  \bibinfo {author} {\bibfnamefont {B.}~\bibnamefont {Delamotte}}, \bibinfo
  {author} {\bibfnamefont {M.}~\bibnamefont {Marohnic}},\ and\ \bibinfo
  {author} {\bibfnamefont {N.}~\bibnamefont {Wschebor}},\ }\bibfield  {title}
  {\bibinfo {title} {{Convergence of Nonperturbative Approximations to the
  Renormalization Group}},\ }\href
  {https://doi.org/10.1103/PhysRevLett.123.240604} {\bibfield  {journal}
  {\bibinfo  {journal} {Phys. Rev. Lett.}\ }\textbf {\bibinfo {volume} {123}},\
  \bibinfo {pages} {240604} (\bibinfo {year} {2019})},\ \Eprint
  {https://arxiv.org/abs/1907.01829} {arXiv:1907.01829 [cond-mat.stat-mech]}
  \BibitemShut {NoStop}%
\bibitem [{Note5()}]{Note5}%
  \BibitemOpen
  \bibinfo {note} {On a lattice, the continuum limit $k^4$ term is replaced by
  a squared lattice Laplacian corresponding to competing nearest and next
  nearest neighbor interactions.}\BibitemShut {Stop}%
\bibitem [{Note6()}]{Note6}%
  \BibitemOpen
  \bibinfo {note} {If higher derivative terms dominate with negative
  coefficients, then even if $Z >0$, it is possible that $E(p)$ about $p=0$ is
  a local minimum, but that the global minimum of $E(p)$ is at $p\protect \neq
  0$.}\BibitemShut {Stop}%
\bibitem [{Note7()}]{Note7}%
  \BibitemOpen
  \bibinfo {note} {This occurs, e.g., for a roton, where $p = 0$ is the global
  minimum, but there is a local minimum for one or more $p_{\protect \rm roton}
  \protect \neq 0$.}\BibitemShut {Stop}%
\bibitem [{\citenamefont {McLerran}\ and\ \citenamefont
  {Toimela}(1985)}]{McLerran:1984ay}%
  \BibitemOpen
  \bibfield  {author} {\bibinfo {author} {\bibfnamefont {L.~D.}\ \bibnamefont
  {McLerran}}\ and\ \bibinfo {author} {\bibfnamefont {T.}~\bibnamefont
  {Toimela}},\ }\bibfield  {title} {\bibinfo {title} {{Photon and Dilepton
  Emission from the Quark - Gluon Plasma: Some General Considerations}},\
  }\href {https://doi.org/10.1103/PhysRevD.31.545} {\bibfield  {journal}
  {\bibinfo  {journal} {Phys. Rev. D}\ }\textbf {\bibinfo {volume} {31}},\
  \bibinfo {pages} {545} (\bibinfo {year} {1985})}\BibitemShut {NoStop}%
\bibitem [{\citenamefont {O'Connell}\ \emph {et~al.}(1997)\citenamefont
  {O'Connell}, \citenamefont {Pearce}, \citenamefont {Thomas},\ and\
  \citenamefont {Williams}}]{OConnell:1995nse}%
  \BibitemOpen
  \bibfield  {author} {\bibinfo {author} {\bibfnamefont {H.~B.}\ \bibnamefont
  {O'Connell}}, \bibinfo {author} {\bibfnamefont {B.~C.}\ \bibnamefont
  {Pearce}}, \bibinfo {author} {\bibfnamefont {A.~W.}\ \bibnamefont {Thomas}},\
  and\ \bibinfo {author} {\bibfnamefont {A.~G.}\ \bibnamefont {Williams}},\
  }\bibfield  {title} {\bibinfo {title} {{$\rho - \omega$ mixing, vector meson
  dominance and the pion form-factor}},\ }\href
  {https://doi.org/10.1016/S0146-6410(97)00044-6} {\bibfield  {journal}
  {\bibinfo  {journal} {Prog. Part. Nucl. Phys.}\ }\textbf {\bibinfo {volume}
  {39}},\ \bibinfo {pages} {201} (\bibinfo {year} {1997})},\ \Eprint
  {https://arxiv.org/abs/hep-ph/9501251} {arXiv:hep-ph/9501251} \BibitemShut
  {NoStop}%
\bibitem [{\citenamefont {Rapp}\ and\ \citenamefont
  {Wambach}(2000)}]{Rapp:1999ej}%
  \BibitemOpen
  \bibfield  {author} {\bibinfo {author} {\bibfnamefont {R.}~\bibnamefont
  {Rapp}}\ and\ \bibinfo {author} {\bibfnamefont {J.}~\bibnamefont {Wambach}},\
  }\bibfield  {title} {\bibinfo {title} {{Chiral symmetry restoration and
  dileptons in relativistic heavy ion collisions}},\ }\href
  {https://doi.org/10.1007/0-306-47101-9_1} {\bibfield  {journal} {\bibinfo
  {journal} {Adv. Nucl. Phys.}\ }\textbf {\bibinfo {volume} {25}},\ \bibinfo
  {pages} {1} (\bibinfo {year} {2000})},\ \Eprint
  {https://arxiv.org/abs/hep-ph/9909229} {arXiv:hep-ph/9909229} \BibitemShut
  {NoStop}%
\bibitem [{\citenamefont {Rapp}\ and\ \citenamefont {van
  Hees}(2016)}]{Rapp:2016xzw}%
  \BibitemOpen
  \bibfield  {author} {\bibinfo {author} {\bibfnamefont {R.}~\bibnamefont
  {Rapp}}\ and\ \bibinfo {author} {\bibfnamefont {H.}~\bibnamefont {van
  Hees}},\ }\bibfield  {title} {\bibinfo {title} {{Thermal Electromagnetic
  Radiation in Heavy-Ion Collisions}},\ }\href
  {https://doi.org/10.1140/epja/i2016-16257-0} {\bibfield  {journal} {\bibinfo
  {journal} {Eur. Phys. J. A}\ }\textbf {\bibinfo {volume} {52}},\ \bibinfo
  {pages} {257} (\bibinfo {year} {2016})},\ \Eprint
  {https://arxiv.org/abs/1608.05279} {arXiv:1608.05279 [hep-ph]} \BibitemShut
  {NoStop}%
\bibitem [{\citenamefont {Rapp}(2024)}]{Rapp:2024grb}%
  \BibitemOpen
  \bibfield  {author} {\bibinfo {author} {\bibfnamefont {R.}~\bibnamefont
  {Rapp}},\ }\bibfield  {title} {\bibinfo {title} {{Electric Conductivity of
  QCD Matter and Dilepton Spectra in Heavy-Ion Collisions}},\ }\href@noop {} {\
   (\bibinfo {year} {2024})},\ \Eprint {https://arxiv.org/abs/2406.14656}
  {arXiv:2406.14656 [hep-ph]} \BibitemShut {NoStop}%
\bibitem [{Note8()}]{Note8}%
  \BibitemOpen
  \bibinfo {note} {This can be derived in a matrix form, where $\Phi = \sigma +
  i \pi ^a \sigma ^a$, with $\sigma ^a$ the Pauli matrices. Under a gauge
  transformation, $\Phi \rightarrow {\protect \rm e}^{i e \theta \sigma ^3}
  \Phi {\protect \rm e}^{- i e \theta \sigma ^3}$, with $D_\mu = \partial _\mu
  - i e A_\mu [ \sigma ^3, ]$. For example, charged pions contribute to the
  usual kinetic term as $|(\partial _\mu - i e A_\mu ) \pi ^+|^2$.}\BibitemShut
  {Stop}%
\bibitem [{Note9()}]{Note9}%
  \BibitemOpen
  \bibinfo {note} {There is also a term which is odd under charge conjugation,
  $\sim F_{0 i} (D_0 \protect \vec {\phi } \cdot D_i \protect \vec {\phi })$.
  Unlike the other electromagnetic couplings, this term couples a photon to
  both charged and neutral (the $\sigma $ and $\pi ^3$) fields. If $\mu
  \protect \neq 0$ implies net electrical charge, then the coefficient of this
  term must vanish at $\mu = 0$.}\BibitemShut {Stop}%
\bibitem [{\citenamefont {Braaten}\ \emph {et~al.}(1990)\citenamefont
  {Braaten}, \citenamefont {Pisarski},\ and\ \citenamefont
  {Yuan}}]{Braaten:1990wp}%
  \BibitemOpen
  \bibfield  {author} {\bibinfo {author} {\bibfnamefont {E.}~\bibnamefont
  {Braaten}}, \bibinfo {author} {\bibfnamefont {R.~D.}\ \bibnamefont
  {Pisarski}},\ and\ \bibinfo {author} {\bibfnamefont {T.-C.}\ \bibnamefont
  {Yuan}},\ }\bibfield  {title} {\bibinfo {title} {{Production of Soft
  Dileptons in the Quark - Gluon Plasma}},\ }\href
  {https://doi.org/10.1103/PhysRevLett.64.2242} {\bibfield  {journal} {\bibinfo
   {journal} {Phys. Rev. Lett.}\ }\textbf {\bibinfo {volume} {64}},\ \bibinfo
  {pages} {2242} (\bibinfo {year} {1990})}\BibitemShut {NoStop}%
\bibitem [{\citenamefont {Rapp}\ \emph {et~al.}(2001)\citenamefont {Rapp},
  \citenamefont {Shuryak},\ and\ \citenamefont {Zahed}}]{Rapp:2000zd}%
  \BibitemOpen
  \bibfield  {author} {\bibinfo {author} {\bibfnamefont {R.}~\bibnamefont
  {Rapp}}, \bibinfo {author} {\bibfnamefont {E.~V.}\ \bibnamefont {Shuryak}},\
  and\ \bibinfo {author} {\bibfnamefont {I.}~\bibnamefont {Zahed}},\ }\bibfield
   {title} {\bibinfo {title} {{A Chiral crystal in cold QCD matter at
  intermediate densities?}},\ }\href
  {https://doi.org/10.1103/PhysRevD.63.034008} {\bibfield  {journal} {\bibinfo
  {journal} {Phys. Rev.}\ }\textbf {\bibinfo {volume} {D63}},\ \bibinfo {pages}
  {034008} (\bibinfo {year} {2001})},\ \Eprint
  {https://arxiv.org/abs/hep-ph/0008207} {arXiv:hep-ph/0008207 [hep-ph]}
  \BibitemShut {NoStop}%
%%CITATION = HEP-PH/0008207;%%
\bibitem [{\citenamefont {Fu}\ \emph {et~al.}(2024{\natexlab{b}})\citenamefont
  {Fu}, \citenamefont {Pawlowski}, \citenamefont {Pisarski}, \citenamefont
  {Rennecke}, \citenamefont {Wen},\ and\ \citenamefont {Yin}}]{frg_moat}%
  \BibitemOpen
  \bibfield  {author} {\bibinfo {author} {\bibfnamefont {W.-J.}\ \bibnamefont
  {Fu}}, \bibinfo {author} {\bibfnamefont {J.~M.}\ \bibnamefont {Pawlowski}},
  \bibinfo {author} {\bibfnamefont {R.~D.}\ \bibnamefont {Pisarski}}, \bibinfo
  {author} {\bibfnamefont {F.}~\bibnamefont {Rennecke}}, \bibinfo {author}
  {\bibfnamefont {R.}~\bibnamefont {Wen}},\ and\ \bibinfo {author}
  {\bibfnamefont {S.}~\bibnamefont {Yin}},\ }\href@noop {} {\bibinfo {title}
  {The moat regime in {QCD}}} (\bibinfo {year} {2024}{\natexlab{b}}),\ \bibinfo
  {note} {work in progress}\BibitemShut {NoStop}%
\bibitem [{\citenamefont {Selke}(1988)}]{Selke}%
  \BibitemOpen
  \bibfield  {author} {\bibinfo {author} {\bibfnamefont {W.}~\bibnamefont
  {Selke}},\ }\bibfield  {title} {\bibinfo {title} {The annni model-
  theoretical analysis and experimental application},\ }\href@noop {}
  {\bibfield  {journal} {\bibinfo  {journal} {Physics Reports}\ }\textbf
  {\bibinfo {volume} {170}},\ \bibinfo {pages} {213} (\bibinfo {year}
  {1988})}\BibitemShut {NoStop}%
\bibitem [{\citenamefont {Aharony}\ and\ \citenamefont {Bruce}(1979)}]{Amnon}%
  \BibitemOpen
  \bibfield  {author} {\bibinfo {author} {\bibfnamefont {A.}~\bibnamefont
  {Aharony}}\ and\ \bibinfo {author} {\bibfnamefont {A.~D.}\ \bibnamefont
  {Bruce}},\ }\bibfield  {title} {\bibinfo {title} {Lifshitz-point critical and
  tricritical behavior in anisotropically stressed perovskites},\ }\href@noop
  {} {\bibfield  {journal} {\bibinfo  {journal} {Phys. Rev. Lett.}\ }\textbf
  {\bibinfo {volume} {42}},\ \bibinfo {pages} {462} (\bibinfo {year}
  {1979})}\BibitemShut {NoStop}%
\bibitem [{\citenamefont {Abrahams}\ and\ \citenamefont
  {Dzyaloshinskii}(1977)}]{Elihu}%
  \BibitemOpen
  \bibfield  {author} {\bibinfo {author} {\bibfnamefont {E.}~\bibnamefont
  {Abrahams}}\ and\ \bibinfo {author} {\bibfnamefont {I.~E.}\ \bibnamefont
  {Dzyaloshinskii}},\ }\bibfield  {title} {\bibinfo {title} {A possible
  lifshitz point for ttf-tcnq},\ }\href@noop {} {\bibfield  {journal} {\bibinfo
   {journal} {Solid State Commun.}\ }\textbf {\bibinfo {volume} {23}},\
  \bibinfo {pages} {883} (\bibinfo {year} {1977})}\BibitemShut {NoStop}%
\bibitem [{\citenamefont {Selinger}(2022)}]{LC2}%
  \BibitemOpen
  \bibfield  {author} {\bibinfo {author} {\bibfnamefont {J.~V.}\ \bibnamefont
  {Selinger}},\ }\bibfield  {title} {\bibinfo {title} {Director deformations,
  geometric frustration, and modulated phases in liquid crystals},\ }\href
  {https://doi.org/10.1146/annurev-conmatphys-031620-105712} {\bibfield
  {journal} {\bibinfo  {journal} {Annual Review of Condensed Matter Physics}\
  }\textbf {\bibinfo {volume} {13}},\ \bibinfo {pages} {49} (\bibinfo {year}
  {2022})}\BibitemShut {NoStop}%
\bibitem [{\citenamefont {Nussinov}\ \emph {et~al.}(2009)\citenamefont
  {Nussinov}, \citenamefont {Vekhter},\ and\ \citenamefont
  {Balatsky}}]{compete_GL}%
  \BibitemOpen
  \bibfield  {author} {\bibinfo {author} {\bibfnamefont {A.}~\bibnamefont
  {Nussinov}}, \bibinfo {author} {\bibfnamefont {I.}~\bibnamefont {Vekhter}},\
  and\ \bibinfo {author} {\bibfnamefont {A.~V.}\ \bibnamefont {Balatsky}},\
  }\bibfield  {title} {\bibinfo {title} {onuniform glassy electronic phases
  from competing local orders},\ }\href@noop {} {\bibfield  {journal} {\bibinfo
   {journal} {Physical Review B}\ }\textbf {\bibinfo {volume} {79}},\ \bibinfo
  {pages} {165122} (\bibinfo {year} {2009})}\BibitemShut {NoStop}%
\bibitem [{\citenamefont {Brazovskii}(1975)}]{Braz}%
  \BibitemOpen
  \bibfield  {author} {\bibinfo {author} {\bibfnamefont {S.~A.}\ \bibnamefont
  {Brazovskii}},\ }\bibfield  {title} {\bibinfo {title} {Phase transition of an
  isotropic system to a nonuniform state},\ }\href@noop {} {\bibfield
  {journal} {\bibinfo  {journal} {Zh. Eksp. Teor. Fiz.}\ }\textbf {\bibinfo
  {volume} {68}},\ \bibinfo {pages} {175} (\bibinfo {year} {1975})}\BibitemShut
  {NoStop}%
\bibitem [{\citenamefont {Brazovskii}\ \emph {et~al.}(1987)\citenamefont
  {Brazovskii}, \citenamefont {Dzyaloshinskii},\ and\ \citenamefont
  {Muratov}}]{Braz2}%
  \BibitemOpen
  \bibfield  {author} {\bibinfo {author} {\bibfnamefont {S.~A.}\ \bibnamefont
  {Brazovskii}}, \bibinfo {author} {\bibfnamefont {I.~E.}\ \bibnamefont
  {Dzyaloshinskii}},\ and\ \bibinfo {author} {\bibfnamefont {A.~R.}\
  \bibnamefont {Muratov}},\ }\bibfield  {title} {\bibinfo {title} {Theory of
  weak crystallization},\ }\href@noop {} {\bibfield  {journal} {\bibinfo
  {journal} {Sov. Phys, JETP}\ }\textbf {\bibinfo {volume} {66}},\ \bibinfo
  {pages} {625} (\bibinfo {year} {1987})}\BibitemShut {NoStop}%
\bibitem [{\citenamefont {Fulde}\ and\ \citenamefont {Ferrell}(1964)}]{FFLO1}%
  \BibitemOpen
  \bibfield  {author} {\bibinfo {author} {\bibfnamefont {P.}~\bibnamefont
  {Fulde}}\ and\ \bibinfo {author} {\bibfnamefont {R.~A.}\ \bibnamefont
  {Ferrell}},\ }\bibfield  {title} {\bibinfo {title} {Superconductivity in a
  strong spin-exchange field},\ }\href
  {https://doi.org/10.1103/PhysRev.135.A550} {\bibfield  {journal} {\bibinfo
  {journal} {Phys. Rev.}\ }\textbf {\bibinfo {volume} {135}},\ \bibinfo {pages}
  {A550} (\bibinfo {year} {1964})}\BibitemShut {NoStop}%
\bibitem [{\citenamefont {Larkin}\ and\ \citenamefont
  {Ovchinnikov}(1964)}]{FFLO2}%
  \BibitemOpen
  \bibfield  {author} {\bibinfo {author} {\bibfnamefont {A.~I.}\ \bibnamefont
  {Larkin}}\ and\ \bibinfo {author} {\bibfnamefont {Y.~N.}\ \bibnamefont
  {Ovchinnikov}},\ }\bibfield  {title} {\bibinfo {title} {{Nonuniform state of
  superconductors}},\ }\href@noop {} {\bibfield  {journal} {\bibinfo  {journal}
  {Zh. Eksp. Teor. Fiz.}\ }\textbf {\bibinfo {volume} {47}},\ \bibinfo {pages}
  {1136} (\bibinfo {year} {1964})}\BibitemShut {NoStop}%
\bibitem [{\citenamefont {Larkin}\ and\ \citenamefont
  {Ovchinnikov}(1965)}]{FFLO3}%
  \BibitemOpen
  \bibfield  {author} {\bibinfo {author} {\bibfnamefont {A.~I.}\ \bibnamefont
  {Larkin}}\ and\ \bibinfo {author} {\bibfnamefont {Y.~N.}\ \bibnamefont
  {Ovchinnikov}},\ }\bibfield  {title} {\bibinfo {title} {{Inhomogeneous state
  of superconductors}},\ }\href@noop {} {\bibfield  {journal} {\bibinfo
  {journal} {Sov. Phys. JETP}\ }\textbf {\bibinfo {volume} {20}},\ \bibinfo
  {pages} {762} (\bibinfo {year} {1965})}\BibitemShut {NoStop}%
\bibitem [{\citenamefont {Lilly}\ \emph {et~al.}(1999)\citenamefont {Lilly},
  \citenamefont {Cooper}, \citenamefont {Eisenstein}, \citenamefont
  {Pfeiffer},\ and\ \citenamefont {West}}]{qhstr2}%
  \BibitemOpen
  \bibfield  {author} {\bibinfo {author} {\bibfnamefont {M.~P.}\ \bibnamefont
  {Lilly}}, \bibinfo {author} {\bibfnamefont {K.~B.}\ \bibnamefont {Cooper}},
  \bibinfo {author} {\bibfnamefont {J.~P.}\ \bibnamefont {Eisenstein}},
  \bibinfo {author} {\bibfnamefont {L.~N.}\ \bibnamefont {Pfeiffer}},\ and\
  \bibinfo {author} {\bibfnamefont {K.~W.}\ \bibnamefont {West}},\ }\bibfield
  {title} {\bibinfo {title} {Evidence for an anisotropic state of
  two-dimensional electrons in high landau levels},\ }\href
  {https://doi.org/10.1103/PhysRevLett.82.394} {\bibfield  {journal} {\bibinfo
  {journal} {Phys. Rev. Lett.}\ }\textbf {\bibinfo {volume} {82}},\ \bibinfo
  {pages} {394} (\bibinfo {year} {1999})}\BibitemShut {NoStop}%
\bibitem [{\citenamefont {Fradkin}\ and\ \citenamefont {Kivelson}(1999)}]{QHE}%
  \BibitemOpen
  \bibfield  {author} {\bibinfo {author} {\bibfnamefont {E.}~\bibnamefont
  {Fradkin}}\ and\ \bibinfo {author} {\bibfnamefont {S.~A.}\ \bibnamefont
  {Kivelson}},\ }\bibfield  {title} {\bibinfo {title} {Liquid-crystal phases of
  quantum hall systems},\ }\href {https://doi.org/10.1103/PhysRevB.59.8065}
  {\bibfield  {journal} {\bibinfo  {journal} {Phys. Rev. B}\ }\textbf {\bibinfo
  {volume} {59}},\ \bibinfo {pages} {8065} (\bibinfo {year}
  {1999})}\BibitemShut {NoStop}%
\bibitem [{\citenamefont {Fogler}\ \emph {et~al.}(1996)\citenamefont {Fogler},
  \citenamefont {Koulakov},\ and\ \citenamefont {Shklovskii}}]{fogler1}%
  \BibitemOpen
  \bibfield  {author} {\bibinfo {author} {\bibfnamefont {M.~M.}\ \bibnamefont
  {Fogler}}, \bibinfo {author} {\bibfnamefont {A.~A.}\ \bibnamefont
  {Koulakov}},\ and\ \bibinfo {author} {\bibfnamefont {B.~I.}\ \bibnamefont
  {Shklovskii}},\ }\bibfield  {title} {\bibinfo {title} {Ground state of a
  two-dimensional electron liquid in a weak magnetic field},\ }\href
  {https://doi.org/10.1103/PhysRevB.54.1853} {\bibfield  {journal} {\bibinfo
  {journal} {Phys. Rev. B}\ }\textbf {\bibinfo {volume} {54}},\ \bibinfo
  {pages} {1853} (\bibinfo {year} {1996})}\BibitemShut {NoStop}%
\bibitem [{\citenamefont {Moessner}\ and\ \citenamefont
  {Chalker}(1996)}]{chalker}%
  \BibitemOpen
  \bibfield  {author} {\bibinfo {author} {\bibfnamefont {R.}~\bibnamefont
  {Moessner}}\ and\ \bibinfo {author} {\bibfnamefont {J.~T.}\ \bibnamefont
  {Chalker}},\ }\bibfield  {title} {\bibinfo {title} {Exact results for
  interacting electrons in high landau levels},\ }\href
  {https://doi.org/10.1103/PhysRevB.54.5006} {\bibfield  {journal} {\bibinfo
  {journal} {Phys. Rev. B}\ }\textbf {\bibinfo {volume} {54}},\ \bibinfo
  {pages} {5006} (\bibinfo {year} {1996})}\BibitemShut {NoStop}%
\bibitem [{\citenamefont {Koulakov}\ \emph {et~al.}(1996)\citenamefont
  {Koulakov}, \citenamefont {Fogler},\ and\ \citenamefont
  {Shklovskii}}]{qhstr1}%
  \BibitemOpen
  \bibfield  {author} {\bibinfo {author} {\bibfnamefont {A.~A.}\ \bibnamefont
  {Koulakov}}, \bibinfo {author} {\bibfnamefont {M.~M.}\ \bibnamefont
  {Fogler}},\ and\ \bibinfo {author} {\bibfnamefont {B.~I.}\ \bibnamefont
  {Shklovskii}},\ }\bibfield  {title} {\bibinfo {title} {Charge density wave in
  two-dimensional electron liquid in weak magnetic field},\ }\href
  {https://doi.org/10.1103/PhysRevLett.76.499} {\bibfield  {journal} {\bibinfo
  {journal} {Phys. Rev. Lett.}\ }\textbf {\bibinfo {volume} {76}},\ \bibinfo
  {pages} {499} (\bibinfo {year} {1996})}\BibitemShut {NoStop}%
\bibitem [{\citenamefont {Du}\ \emph {et~al.}(1999)\citenamefont {Du},
  \citenamefont {Tsui}, \citenamefont {Stormer}, \citenamefont {Pfeiffer},
  \citenamefont {Baldwin},\ and\ \citenamefont {West}}]{qhstr3}%
  \BibitemOpen
  \bibfield  {author} {\bibinfo {author} {\bibfnamefont {R.}~\bibnamefont
  {Du}}, \bibinfo {author} {\bibfnamefont {D.}~\bibnamefont {Tsui}}, \bibinfo
  {author} {\bibfnamefont {H.}~\bibnamefont {Stormer}}, \bibinfo {author}
  {\bibfnamefont {L.}~\bibnamefont {Pfeiffer}}, \bibinfo {author}
  {\bibfnamefont {K.}~\bibnamefont {Baldwin}},\ and\ \bibinfo {author}
  {\bibfnamefont {K.}~\bibnamefont {West}},\ }\bibfield  {title} {\bibinfo
  {title} {Strongly anisotropic transport in higher two-dimensional landau
  levels},\ }\href
  {https://doi.org/https://doi.org/10.1016/S0038-1098(98)00578-X} {\bibfield
  {journal} {\bibinfo  {journal} {Solid State Communications}\ }\textbf
  {\bibinfo {volume} {109}},\ \bibinfo {pages} {389} (\bibinfo {year}
  {1999})}\BibitemShut {NoStop}%
\bibitem [{\citenamefont {Kapusta}\ and\ \citenamefont
  {Toimela}(1988)}]{Kapusta:1988fi}%
  \BibitemOpen
  \bibfield  {author} {\bibinfo {author} {\bibfnamefont {J.~I.}\ \bibnamefont
  {Kapusta}}\ and\ \bibinfo {author} {\bibfnamefont {T.}~\bibnamefont
  {Toimela}},\ }\bibfield  {title} {\bibinfo {title} {{Friedel Oscillations in
  Relativistic {QED} and {QCD}}},\ }\href
  {https://doi.org/10.1103/PhysRevD.37.3731} {\bibfield  {journal} {\bibinfo
  {journal} {Phys. Rev. D}\ }\textbf {\bibinfo {volume} {37}},\ \bibinfo
  {pages} {3731} (\bibinfo {year} {1988})}\BibitemShut {NoStop}%
\bibitem [{\citenamefont {Gasser}\ and\ \citenamefont
  {Leutwyler}(1984)}]{Gasser:1983yg}%
  \BibitemOpen
  \bibfield  {author} {\bibinfo {author} {\bibfnamefont {J.}~\bibnamefont
  {Gasser}}\ and\ \bibinfo {author} {\bibfnamefont {H.}~\bibnamefont
  {Leutwyler}},\ }\bibfield  {title} {\bibinfo {title} {{Chiral Perturbation
  Theory to One Loop}},\ }\href {https://doi.org/10.1016/0003-4916(84)90242-2}
  {\bibfield  {journal} {\bibinfo  {journal} {Annals Phys.}\ }\textbf {\bibinfo
  {volume} {158}},\ \bibinfo {pages} {142} (\bibinfo {year}
  {1984})}\BibitemShut {NoStop}%
\bibitem [{\citenamefont {Gasser}\ and\ \citenamefont
  {Leutwyler}(1985)}]{Gasser:1984gg}%
  \BibitemOpen
  \bibfield  {author} {\bibinfo {author} {\bibfnamefont {J.}~\bibnamefont
  {Gasser}}\ and\ \bibinfo {author} {\bibfnamefont {H.}~\bibnamefont
  {Leutwyler}},\ }\bibfield  {title} {\bibinfo {title} {{Chiral Perturbation
  Theory: Expansions in the Mass of the Strange Quark}},\ }\href
  {https://doi.org/10.1016/0550-3213(85)90492-4} {\bibfield  {journal}
  {\bibinfo  {journal} {Nucl. Phys. B}\ }\textbf {\bibinfo {volume} {250}},\
  \bibinfo {pages} {465} (\bibinfo {year} {1985})}\BibitemShut {NoStop}%
\bibitem [{\citenamefont {Manohar}(2018)}]{Manohar:2018aog}%
  \BibitemOpen
  \bibfield  {author} {\bibinfo {author} {\bibfnamefont {A.~V.}\ \bibnamefont
  {Manohar}},\ }\href {https://doi.org/10.1093/oso/9780198855743.003.0002}
  {\emph {\bibinfo {title} {{Introduction to Effective Field Theories}}}},\
  edited by\ \bibinfo {editor} {\bibfnamefont {S.}~\bibnamefont {Davidson}},
  \bibinfo {editor} {\bibfnamefont {P.}~\bibnamefont {Gambino}}, \bibinfo
  {editor} {\bibfnamefont {M.}~\bibnamefont {Laine}}, \bibinfo {editor}
  {\bibfnamefont {M.}~\bibnamefont {Neubert}},\ and\ \bibinfo {editor}
  {\bibfnamefont {C.}~\bibnamefont {Salomon}}\ (\bibinfo {year} {2018})\
  \Eprint {https://arxiv.org/abs/1804.05863} {arXiv:1804.05863 [hep-ph]}
  \BibitemShut {NoStop}%
\bibitem [{\citenamefont {Stewart}(2014)}]{Iain-EFT}%
  \BibitemOpen
  \bibfield  {author} {\bibinfo {author} {\bibfnamefont {I.~W.}\ \bibnamefont
  {Stewart}},\ }\bibfield  {title} {\bibinfo {title}
  {\href{https://courses.edx.org/c4x/MITx/8.EFTx/asset/notes_EFT.pdf}{Effective
  Field Theory Lecture Notes}},\ }\href@noop {} {\bibfield  {journal} {\bibinfo
   {journal} {EdX}\ } (\bibinfo {year} {2014})}\BibitemShut {NoStop}%
\bibitem [{\citenamefont {Gross}\ \emph {et~al.}(2023)\citenamefont {Gross}
  \emph {et~al.}}]{Gross:2022hyw}%
  \BibitemOpen
  \bibfield  {author} {\bibinfo {author} {\bibfnamefont {F.}~\bibnamefont
  {Gross}} \emph {et~al.},\ }\bibfield  {title} {\bibinfo {title} {{50 Years of
  Quantum Chromodynamics}},\ }\href
  {https://doi.org/10.1140/epjc/s10052-023-11949-2} {\bibfield  {journal}
  {\bibinfo  {journal} {Eur. Phys. J. C}\ }\textbf {\bibinfo {volume} {83}},\
  \bibinfo {pages} {1125} (\bibinfo {year} {2023})},\ \Eprint
  {https://arxiv.org/abs/2212.11107} {arXiv:2212.11107 [hep-ph]} \BibitemShut
  {NoStop}%
\bibitem [{\citenamefont {Arzt}(1995)}]{Arzt:1993gz}%
  \BibitemOpen
  \bibfield  {author} {\bibinfo {author} {\bibfnamefont {C.}~\bibnamefont
  {Arzt}},\ }\bibfield  {title} {\bibinfo {title} {{Reduced effective
  Lagrangians}},\ }\href {https://doi.org/10.1016/0370-2693(94)01419-D}
  {\bibfield  {journal} {\bibinfo  {journal} {Phys. Lett. B}\ }\textbf
  {\bibinfo {volume} {342}},\ \bibinfo {pages} {189} (\bibinfo {year}
  {1995})},\ \Eprint {https://arxiv.org/abs/hep-ph/9304230}
  {arXiv:hep-ph/9304230} \BibitemShut {NoStop}%
\end{thebibliography}%
\clearpage
\appendix
\onecolumngrid
\section*{Supplemental material}
\subsection{Spurious van Hove singularities from gauge non-invariance}
\label{app:supplementalMaterial}

Here we compare our analysis to that of Reference~\cite{Hayashi:2024sae}, which appeared while this work was in preparation. 
First, we note that the target of our analyses differ, in that Reference~\cite{Hayashi:2024sae} 
studied dilepton production in the presence of 
one specific inhomogeneous field configuration: 
a single chiral spiral.  While chiral spirals {\it can}
appear in a moat regime, it is generally not necessary
that they {\it do} appear; indeed, a moat regime need not give
rise to \textit{any} spatially inhomogeneous phase. 
Reference \cite{Hayashi:2024sae} 
did not use a gauge invariant form for the higher order derivative terms. Thus their 
$\Gamma^\mu$ does not satisfy the Ward identity of Eq. (\ref{ward_identity}), and the associated photon self energy is not transverse. 
Lastly, the new gauge invariant operators to sixth order in the mass dimension, Eqs. (\ref{der_ints1})-(\ref{der_ints4}), are novel to our analysis.

\begin{figure*}[ht]
    \centering
   \includegraphics{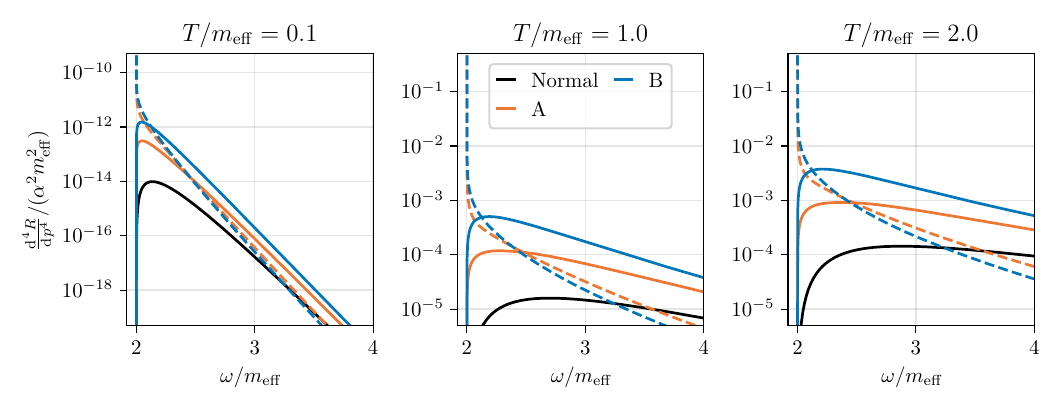}
   \caption{
   Dilepton production rate in Eq.~\eqref{eq:dilepton-production} as a function of $\omega$ at $T/m_\mathrm{eff}=0.1,0.5,1.0$, for the dispersion relations in Fig. \ref{fig:dispersionAndDilepton}, at $\vec p=0$. Solid lines represent the correct result; dashed lines are the result where only the quadratic term is gauged.  }
   \label{fig:gauge}
\end{figure*}

As discussed in the main body of the manuscript, the neglect of higher derivative terms in the vertex leads to the appearance of a van Hove singularity.
To illustrate how the van Hove singularity is smoothed out by
correctly imposing gauge invariance, in Fig.~\ref{fig:gauge} we compute dilepton production in two different ways. 
Both use Eq.~\eqref{eq:dilepton-production} 
at $p=0$, and the dispersion relations in Fig. \ref{fig:dispersionAndDilepton}.
Solid lines represent the correct result, where both the quadratic and quartic terms are gauged; dashed lines are the result where only the quadratic term is gauged. 
For the former, the threshold behavior of pions with a moat is
smooth, like normal pions; for the latter, there are
van Hove singularities, which are thus unphysical at $p=0$.
Neglecting gauge invariance for higher order terms is equivalent to taking $M \rightarrow  \infty$ for the photon-$\phi$-$\phi$ vertex, $\Gamma^\mu$.  
Such a vertex does not satisfy the Ward identity of
Eq.~(\ref{ward_identity}), so the photon
self energy, $\Pi^{\mu \nu}$ is not
transverse, $P^\mu \Pi^{\mu \nu} \neq 0$. 
As described in the main part of the manuscript, the correct vertex structure near threshold, Eq.~\eqref{vertex_canc}, smooths out the van Hove singularity, and leads to the observed enhancement of dilepton production.
In Ref. \cite{Hayashi:2024sae}, the integrals over phase space
were modified to eliminate the van Hove singularities.

\subsection{Relationship with chiral perturbation theory }

In the main text, we write down a dilepton effective Lagrangian for sigmas and pions, which might prompt one to wonder about the relation between our effective Lagrangian
and other effective theories of light mesons, such as chiral perturbation theory ($\chi$PT) \cite{Gasser:1983yg,Gasser:1984gg}.\\

\noindent{\bf Chiral perturbation theory.} Chiral perturbation theory is a bottom-up effective field theory (EFT) built from light meson fields that respect the global symmetries of QCD and its broken chiral symmetry SU($N_f$)$_L \times$ SU($N_f$)$_R\to $SU($N_f$)$_V$. In the case of $N_f=2$, these are the pions, and in the case of $N_f=3$, we also include kaons and etas.
At lowest order, the $\chi$PT Lagrangian has two terms:
\begin{align}\label{eq:chipt-o2}
{\cal L}_{\chi PT}^{(p^2)}
=\frac{F^2_0}{4}\Tr (\partial_\mu U \partial^\mu U^\dagger)
+\Tr(\chi^\dagger U+U^\dagger\chi)
\end{align}
Here, $U = \exp(i\phi F_0)$ encodes the meson fields in $\phi$ and $\chi$ encodes their masses, which break chiral symmetry.
The pion decay constant $F_0\approx 93$ MeV normalizes the Lagrangian to have the typical kinetic term in $\phi$, as seen from the expansion $U = 1 + i\phi/F_0 + ...\, $.
These fields power count as $U\sim p^0$, $D_\mu U \sim p^1$, $\chi \sim p^2$.

We can write down the most general possible linearly independent set of $\cO(p^4)$ terms of the $\chi$PT Lagrangian as
\begin{align}\label{eq:chipt-o4}
	\cL_{\chi PT}^{(p^4)}& = L_1 \left\{\Tr[D_{\mu}U (D^{\mu}U)^{\dagger}] \right\}^2
%    \nonumber\\
	+ L_2 \,\Tr [D_{\mu}U (D_{\nu}U)^{\dagger}]
	\Tr [D^{\mu}U (D^{\nu}U)^{\dagger}]
 %   \nonumber\\
	+ L_3 \,\Tr[
	D_{\mu}U (D^{\mu}U)^{\dagger}D_{\nu}U (D^{\nu}U)^{\dagger}]
    \nonumber\\
     &+ L_4\, \Tr [ D_{\mu}U (D^{\mu}U)^{\dagger} ]
	\Tr ( \chi U^{\dagger}+ U \chi^{\dagger} )
%   \nonumber\\
	+ L_5 \,\Tr [ D_{\mu}U (D^{\mu}U)^{\dagger}
	(\chi U^{\dagger}+ U \chi^{\dagger})]
 %   \nonumber\\
	+ L_6 [\Tr ( \chi U^{\dagger}+ U \chi^{\dagger}  )
	]^2
    \nonumber\\
    & + L_7\,[ \Tr ( \chi U^{\dagger} - U \chi^{\dagger})]^2
%    \nonumber\\
	 + L_8 \Tr ( U \chi^{\dagger} U \chi^{\dagger}
	+ \chi U^{\dagger} \chi U^{\dagger})
%    \nonumber\\
	 -i L_9 \Tr[ f^R_{\mu\nu} D^{\mu} U (D^{\nu} U)^{\dagger}+ f^L_{\mu\nu} (D^{\mu} U)^{\dagger} D^{\nu} U ]
    \nonumber\\
	& + L_{10}\,\Tr ( U f^L_{\mu\nu} U^{\dagger} f_R^{\mu\nu})\,,
\end{align}
where $L_{1-10}$ are dimensionless constants encoding higher-energy information about QCD that is not described explicitly in $\chi$PT.\\

\noindent \textbf{Bottom-up effective field theory.} 
A rigorous EFT is a description of a system that (1) takes the form of a power expansion, which (2) includes \textit{all} possible relevant operators at a given order that satisfy \textit{all} symmetry conditions imposed on the system \cite{Manohar:2018aog, Iain-EFT, Gross:2022hyw}. An EFT is thus systematically improvable, to an arbitrary degree of precision. There are two types of EFTs: top-down and bottom-up. A top-down EFT is constructed by integrating modes out of $\cL_{\rm QCD}$ systematically, whereas a bottom-up EFT is built from a general operator basis without any assumptions on the nature of the high-energy Lagrangian. We construct a bottom-up EFT from a set of building block fields by writing down the most general set of operators $O_j^{(i)}$ that carry a given power counting (scaling) at each order $\cO(\lambda^i)$. Schematically,
\begin{align}
	&\cL_{\rm EFT} \equiv \cL_{\rm EFT}^{(0)} + \cL_{\rm EFT}^{(1)} + \cL_{\rm EFT}^{(2)} + ...\,,
	&&\cL^{(i)}_{\rm EFT} = \sum_{j} c_j^{(i)}O_j^{(i)}\,.
\end{align}
Here, we determine the couplings $c_j^{(i)}$ by fitting to experiment. 
In EFT, we can generally whittle down the general set of possible $\cO(\lambda^i)$ operators down to a complete, linearly independent set by using the equations of motion \cite{Arzt:1993gz}. It is in general true that \textit{no} new operators must be added at $\cO(\lambda^n)$ when we construct next-order terms of the Lagrangian at $\cO(\lambda^{i+1})$. 
One key difference between a rigorous bottom-up EFT and a bottom-up model is that the EFT considers every single operator possible given these constraints, and is thus uniquely determined, whereas a model includes only some subset of the possible operators.\\

\noindent{\bf Comparison between $\chi$PT and the dilepton effective Lagrangian.}

Let us now compare eq.~\eqref{eq:chipt-o4} to our dilepton Lagrangian in the main text. 
First, it is immediately clear that $\chi$PT and the dilepton Lagrangian are both constructed in the style of a bottom-up effective field theory; however, $\chi$PT is a rigorous EFT of QCD but the dilepton Lagrangian is not. 
An important feature of $\chi$PT is that it is constructed at zero density, and thus describes a Hermitian and Lorentz-invariant system. Thus, we know that we can truncate $\chi$PT at a given order when carrying out precision calculations for collliders and get reasonable results. In contrast, the dilepton Lagrangian is for a non-Hermitian and non-Lorentz-invariant system. When truncating the dilepton expansion at any given order in mass dimension, we have to worry that negative couplings could appear at a higher order than our truncation, and thus that we could miss important effects like additional local minima in the dispersion relation. Writing down the $-Z$ term allows us to  capture the key features of moatons, but is not a systematically improvable choice that could be used in precision studies. Thus we refer to the dilepton Lagrangian as merely an effective Lagrangian and \textit{not} the proper EFT for QCD in a moat regime.
 
Second, we consider the phenomenology. $\chi$PT is an EFT describing the interactions of light mesons at zero density. This contrasts from the models for dilepton production in the main text, which are primarily concerned with a subset of meson fields $\vec{\phi}$ chosen on physical grounds and their effect on the photon field $A_\mu$. Moreover, in the dilepton production case, we are analyzing a dense baryonic system, so we must consider time derivatives and spatial derivatives separately to achieve the full set of linearly independent terms.

Third, let us compare the structure of the two Lagrangians. Although the underlying phenomenology of the two situations is slightly different, we expect that there is a direct relationship between their structure at each order in the power counting. In both cases, we use a set of building block operators that have a parallel power counting structure. Namely, we can draw a comparison $\vec \phi \sim U\sim p^0$, $D_\mu \vec\phi \sim D_\mu U \sim p^1$, and $F_{\mu\nu} \sim f_{\mu\nu}^{L/R}\sim p^2$. On simple algebraic grounds, from two similar sets of building blocks we should construct two similar operator bases for the effective Lagrangian at any given order $\cO(p^{n})$.

Let us see this intuition in action. Comparing the dilepton and $\chi$PT Lagrangians, we immediately notice that we have no equivalent of the $\chi$ field in the dilepton case, and thus we can ignore the terms $L_4$-$L_8$ in eq.~\eqref{eq:chipt-o4}. Next, we recall that the dilepton Lagrangian is at finite density but eq.~\eqref{eq:chipt-o4} describes the behavior of mesons in a vacuum. Thus, a meaningful comparison requires us to take the zero-density limit of the dilepton Lagrangian. In practice, this requires us to set $Z=1$ and combine dilepton terms in which the  time and spatial derivatives are separated. Combining such terms fixes $C_1 = C_2$, $C_3=C_4$, $C_5=C_6$, 
and $C_7=C_8$, 
so that
\begin{align}\label{eq:dilepton-reduce}
    &\cL_{\rm dilepton}^{(p^2,\mu=0)} = \frac{1}{2} (D_\mu \vec{\phi})^2 +  
    \frac{C_1}{M^2} (D_\mu \vec\phi)^2 |\vec\phi|^2
    + \frac{C_3}{M^2} (\vec{\phi}\cdot D_\mu \vec\phi)^2  
    \,,
    \nonumber\\
    &\cL_{\rm dilepton}^{(p^4,\mu=0)} = 
    \frac{1}{2 M^2} (D_\mu^2 \vec{\phi})^2 
 %   \nonumber\\
    + \frac{C_5}{M^2}F_{\mu\nu}^2 |\vec\phi|^2 
  %  \nonumber\\
    + \frac{e C_7}{M^2} F_{\mu\nu} (D_\mu\vec{\phi} \cdot
    i t_2 \cdot D_\nu \vec{\phi})\,.
\end{align}
We immediately see that $C_1$ and $C_3$ align with our $\cO(p^2)$ $\chi$PT Lagrangian. Additionally, two pairs of terms in the $\cO(p^4)$ terms of eqs.~\eqref{eq:dilepton-reduce} and \eqref{eq:chipt-o4} take on similar forms: $(C_5, L_{10})$ and $(C_7, L_9)$.
The terms $L_{1-3}$ contain 4 powers of $D_\mu$, which are reminiscent of the term $(D_\mu^2 \vec{\phi})^2/(2 M^2)$; however, $L_{1-3}$ do not contain an explicit $\Tr[D^2U D^2 U^\dagger]$. Nonetheless, the structure we need can be obtained from these terms by using the equations of motion; specifically, ignoring the $\chi$ terms as previously in $\chi$PT, we have that $  (\partial^2U^\dagger)(\partial^2U)=(\partial_\mu U^\dagger \partial_\mu U)^2$, allowing us to identify $L_1$ with the first term of $\cL_{\rm dilepton}^{(p^4,\mu=0)}$. The terms $L_2$ and $L_3$ do not contribute to the imaginary part of the photon self-energy at lowest order. 
\end{document}